\documentclass[%
 reprint,twocolumn,aps,superscriptaddress,
 amsmath,amssymb,
 aps,
floatfix,longbibliography
]{revtex4-2}
\usepackage{color} 
\usepackage{graphicx}
\usepackage{dcolumn}
\usepackage{bm}
\usepackage{physics, bm,bbm, dsfont}
\usepackage{mathrsfs}  
\usepackage{amsmath, amssymb, amscd, amsthm, amsfonts}
\usepackage{xparse}
\usepackage{babel}[english]
\usepackage{soul}

\newcommand{\bxi}{\boldsymbol{\xi}}

\newcommand{\bGamma}{\boldsymbol{\Gamma}}

\newcommand{\beeta}{\boldsymbol{\eta}}

\newcommand{\p}{\partial}
\newcommand{\bx}{\text{\bf x}}
\newcommand{\br}{\text{\bf r}}

\newcommand{\bu}{\text{\bf u}}
\newcommand{\bR}{\text{\bf R}}

\newcommand{\bnabla}{\boldsymbol{\nabla}}
\newcommand{\bmu}{\boldsymbol{\mu}}

\newcommand{\by}{\text{\bf y}}

\newcommand{\bF}{\text{\bf F}}
\newcommand{\bA}{\text{\bf A}}
\newcommand{\bXi}{\boldsymbol{\Xi}}
\newcommand{\bM}{\text{\bf M}}
\newcommand{\bD}{\text{\bf D}}
\newcommand{\brhat}{\boldsymbol{\hat{\textbf{r}}}}
\newcommand{\bnu}{\boldsymbol{\nu}}
\newcommand{\bpartial}{\boldsymbol{\partial}}
\newcommand{\bFext}{\text{\bf F}^\text{ext}}
\newcommand{\Lext}{L^{\text{ext}}}
\newcommand{\mPiort}[1]{\boldsymbol{\Pi}_{\perp}(#1)}

\ExplSyntaxOn
\NewDocumentCommand{\eqrefs}{m}
 {
  \joansola_eqrefs:n { #1 }
 }

\seq_new:N \l__joansola_eqrefs_in_seq
\seq_new:N \l__joansola_eqrefs_out_seq

\cs_new_protected:Nn \joansola_eqrefs:n
 {
  \seq_set_split:Nnn \l__joansola_eqrefs_in_seq { , } { #1 }
  \seq_set_map:NNn \l__joansola_eqrefs_out_seq \l__joansola_eqrefs_in_seq
   {
    \exp_not:n { \ref{##1} }
   }
  \int_compare:nTF { \seq_count:N \l__joansola_eqrefs_out_seq < 2 }
   { Eq.\nobreakspace }
   { Eqs.\nobreakspace }
  \textup{(\seq_use:Nn \l__joansola_eqrefs_out_seq {,~})}
 }
\ExplSyntaxOff

\begin{document}

\title{Transverse forces and glassy liquids in infinite dimensions}

\author{Federico Ghimenti}
\affiliation{Laboratoire Mati\`ere et Syst\`emes Complexes (MSC), Université Paris Cité  \& CNRS (UMR 7057), 75013 Paris, France}

\author{Ludovic Berthier}
\affiliation{Laboratoire Charles Coulomb (L2C), Université de Montpellier \& CNRS (UMR 5221), 34095 Montpellier, France}
\affiliation{Gulliver, UMR CNRS 7083, ESPCI Paris, PSL Research University, 75005 Paris, France}

\author{Grzegorz Szamel}
\affiliation{Department of Chemistry, Colorado State University, Fort Collins, Colorado 80523, United States of America}

\author{Fr\'ed\'eric van Wijland}
\affiliation{Laboratoire Mati\`ere et Syst\`emes Complexes (MSC), Université Paris Cité  \& CNRS (UMR 7057), 75013 Paris, France}

\date{\today}

\begin{abstract}
We explore the dynamics of a simple liquid whose particles, in addition to standard potential-based interactions, are also subjected to transverse forces  preserving the Boltzmann distribution. We derive the effective dynamics of one and two tracer particles in the infinite-dimensional limit. We determine the amount of acceleration of the dynamics caused by the transverse forces, in particular in the vicinity of the glass transition. We analyze the emergence and evolution of odd transport phenomena induced by the transverse forces. 
\end{abstract}

\maketitle

\section{Introduction}

\label{intro}

As temperature is decreased, the relaxation of a glass-forming liquid takes an increasingly longer time to unfold~\cite{ediger1996supercooled,berthier2011theoretical}. Many liquids actually undergo an impressive dynamical slowing down, with the viscosity and the structural relaxation time increasing by many orders of magnitude upon a mild decrease of the temperature. To explore the dynamics of such a liquid on a computer, the system has to be brought into an equilibrium state before its response or relaxation properties can be investigated~\cite{berthier2023modern}. It may happen that simply reaching this initial equilibrium state is a difficult problem, especially at low temperatures or high densities. It is thus a challenge on its own to correctly sample the Boltzmann distribution. A variety of methods have been used to reach equilibrium in the most efficient manner~\cite{berthier2023modern,barrat2023computer}. A very efficient approach is the swap Monte Carlo algorithm~\cite{grigera2001fast,berthier2016equilibrium,ninarello2017models,berthier2019efficient}, which relies upon unphysical radius exchange moves that are performed respecting the detailed balance condition. When applied to size polydisperse mixtures this technique dramatically reduces the equilibration time by several orders of magnitude~\cite{ninarello2017models}. 

Because there is no need to respect realistic physics during the equilibration phase, one could utilize detailed balance violating, \textit{i.e} nonequilibrium, dynamics, as long as sampling  the Boltzmann distribution is guaranteed in the steady-state~\cite{krauth2006statistical}. This idea of resorting to  tailored nonequilibrium dynamics originated in the applied mathematical literature: it consists in inducing an extra current in the system that, while driving the system out of equilibrium, maintains the Boltzmann distribution in the steady state. The potential equilibration speedup results from a theorem that explains the conditions under which a nonequilibrium drive makes the relaxation times shorter. This was proven for the first time in Ref.~\cite{hwang1993accelerating} for overdamped Langevin dynamics in a harmonic potential, and it was then extended to general confining potentials~\cite{hwang2005accelerating} and discrete Markov chains~\cite{ichiki2013violation}. 

There are many ways by which the out-of-equilibrium drive can be implemented, but two main families appear in the literature. A first one~\cite{hwang1993accelerating} consists in applying to the degrees of freedom of the system a transverse force, perpendicular to the direction of the energy gradient. In the second one, called lifting, one increases the number of degrees of freedom of the system, exploiting the extended phase space to induce the nonequilibrium current~\cite{vucelja2016lifting,krauth2021event}. Inequalities on relaxation and mixing times \cite{chen1999lifting, diaconis2000analysis} can be proven for both dynamical evolutions, ensuring faster equilibration compared to the detailed balance satisfying dynamics. However, the question of how these approaches perform in a challenging sampling problem, as for instance the equilibration of systems characterized by rough energy landscapes, is still largely open. In what follows, we restrict our analysis to transverse forces, because they represent the minimal nonequilibrium ingredient to achieve acceleration, thus making the comparison with equilibrium dynamics easier. This task is more complicated for lifted dynamics, where the nonequilibrium drive is not simply added to the original equilibrium evolution, making it difficult to understand how much of the speedup is due to the out-of-equilibrium nature of the dynamics. 

The physics of dense liquids remains a puzzle because the microscopic mechanisms controlling the dynamics of the liquid, as well as the interplay between dynamical evolution and structural changes, are not fully understood in finite dimensional systems~\cite{scalliet2022thirty,bouchaud2024dynamics}. On the other hand, in the infinite-dimensional limit, a more complete level of understanding is now available~\cite{charbonneau2017glass,parisi2020theory}. For liquids, the mean field condition can be implemented \cite{frisch1985classical,kurchan2012exact} by sending to infinity the number of spatial dimensions of the system, while properly rescaling the number density so that the average number of neighbors per particle grows linearly with the dimension. In this limit, the dynamics of the system can be analytically studied~\cite{maimbourg2016solution, agoritsas2019out1, agoritsas2019out2, liu2021dynamics, biroli2022local}, and expressions for transport coefficients such as the diffusion constant and the viscosity can be obtained. Below a dynamical transition temperature, $T_d$, the diffusion constant vanishes, signalling a dynamical ergodicity breaking. In mean-field this dynamical transition can be inferred from a study of the thermodynamics of the system~\cite{mezard1999thermodynamics, parisi2010mean, parisi2020theory}, as it corresponds to the temperature below which infinitely long-lived metastable glassy states appear~\cite{monasson1995structural,franz1997phase}.

The theoretical framework developed to obtain the aforementioned results makes liquids in high dimension a suitable testbench for the question we ask, namely how efficiently a given dynamics samples the Boltzmann distribution in a high dimensional rugged landscape. This challenging task is applicable for a wide set of topics going beyond structural glasses, ranging from protein folding to machine learning algorithms~\cite{futami2020accelerating, gao2020breaking}. 

\begin{figure}
  \includegraphics[width = \columnwidth]{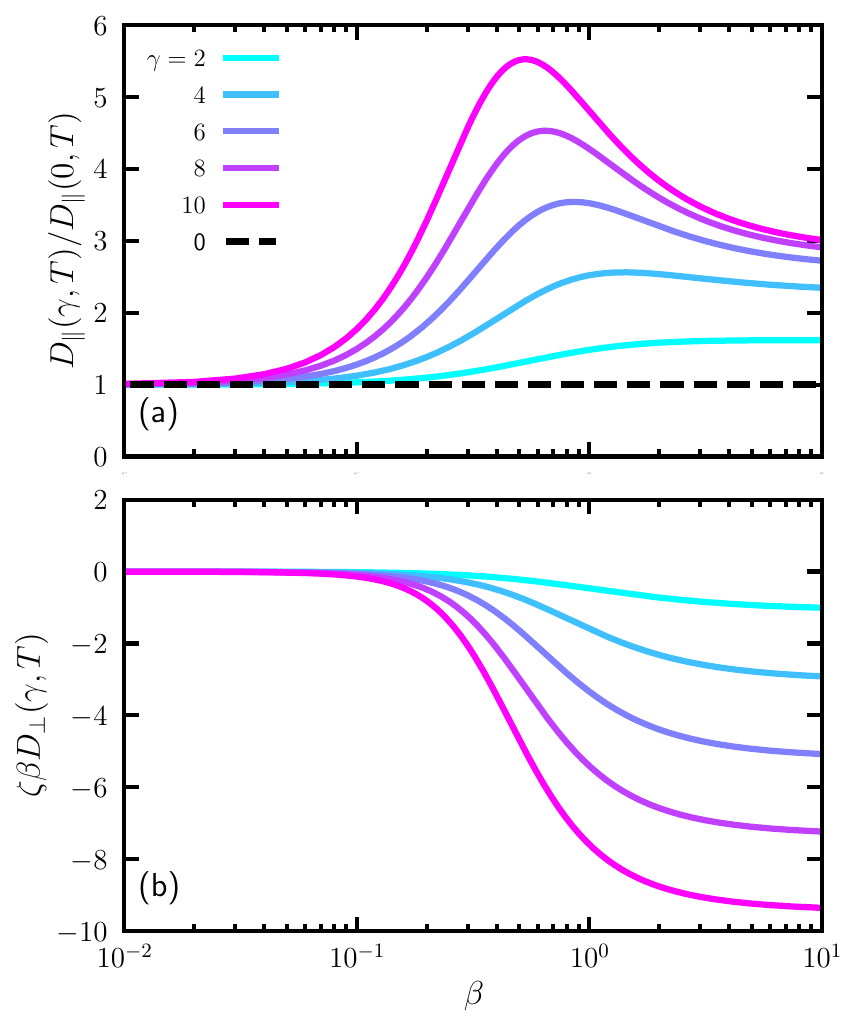}
  \caption{(a) Ratio between the longitudinal diffusion constant $D_\parallel(\gamma, T)$ in the presence of transverse forces and its equilibrium counterpart at $\gamma=0$ for different values $\gamma$ of the strength of the nonequilibrium drive, as a function of inverse temperature $\beta$. (b) Odd diffusivity in the presence of transverse forces. In  both panels, the memory kernel used is the one obtained via a low density expansion for the case of a linear potential, as given in Eq.~\eqref{eq:M_lowphi}.}
  \label{fig:Drel_lowphi}
\end{figure}

In the present article, we evaluate the efficiency of transverse forces in liquids living in infinite-dimensions, explaining and expanding on results announced in a recent Letter~\cite{ghimenti2023sampling}. We use the diffusion constant to probe the relaxation dynamics and to quantify the effect of transverse forces. We analyze the resulting speedup as a function of the temperature and we elucidate the scaling of the acceleration for large values of the nonequilibrium drive. We find that after reaching a maximum in the ergodic region, the efficiency decreases towards a constant value as the dynamical glass transition is approached, as shown in Fig.~\ref{fig:Drel_lowphi}. This is the main result of this work. The temperature $T_d$ at which the dynamical glass transition occurs is unaffected by the presence of transverse forces, suggesting that the long lived, metastable states that hinder the relaxation for this particular dynamics are encapsulated in the equilibrium distribution, which is preserved by transverse forces~\cite{ghimenti2022accelerating}. This is already an interesting result in itself, since {\it a priori} $T_d$ may depend on the specifics of the dynamics, as discussed in \cite{ikeda2017mean, szamel2018theory}.

Our study elucidates the role of the transverse forces in the relaxation speedup. We pinpoint the influence of transverse forces in the one and two particle processes characterizing the dynamics of mean-field liquids. The technical difficulty is to carefully control the influence  of the forces perpendicular to the energy gradient, which we achieve by extending the cavity approach of Ref.~\cite{liu2021dynamics}. On the physics side, transverse forces give rise to odd transport coefficients (diffusivity~\cite{hargus2021odd}, mobility~\cite{poggioli2023odd}, and  viscosity~\cite{banerjee2017odd}), which also arise naturally in active systems composed of chiral particles or in the presence of nonreciprocal forces~\cite{fruchart2021non}. These coefficients usefully quantify the departure from equilibrium dynamics and explain how the relaxational dynamics is driven by nonequilibrium currents even in cases where the speedup efficiency is reduced.

This paper is organized as follows. In Sec.~\ref{sec:eom} we introduce the equations of motion and discuss both transverse forces and the proper infinite dimensional scaling of parameters. In Secs.~\ref{sec:1p} and \ref{sec:2p} we study respectively the one and two-particle processes. From the asymptotic properties of the latter, the temperature at which dynamical arrest occurs can be predicted. This is done in Sec.~\ref{sec:dynarr}. We discuss the dynamics in the ergodic regime for large values of the out-of-equilibrium driving in Sec.~\ref{sec:gammalarge}. We compute the mean squared displacement and discuss transverse force efficiency in Sec.~\ref{sec:msd}. Finally, in Secs.~\ref{sec:odddiff},~\ref{sec:oddmob} and~\ref{sec:oddvisc} we respectively compute the odd diffusion constant, the odd mobility, and the odd viscosity.

\section{Equations of motion}

\label{sec:eom}

Our starting point is an overdamped Langevin dynamics in the presence of transverse forces for a system of interacting particles with positions $\bR_i$ in a space of dimension $d$:
\begin{equation}
\zeta\dot\bR_i = -(\mathds{1}_d+\gamma\bA)\bnabla U + \bxi(t) ,
\end{equation}
where $U= \sum_{i<j}v(\bR_i(t) - \bR_j(t))$, with potential $v$ being pairwise and spherically symmetric; $\bxi$ is a Gaussian white noise with correlations $\left\langle \bxi(t) \otimes \bxi(t')\right\rangle = 2\zeta T\mathds{1}_d \delta(t-t')$, where $\zeta$ is the friction coefficient. As in the equilibrium case examined in \cite{agoritsas2019out1}, we will focus on the dynamics of the displacements $\bu_i(t) \equiv \bR_i(t) - \bR_i(0)$ with respect to the initial positions at $t=0$. We impose that the Frobenius norm of $\mathbf A$ is $||{\mathbf A}||_\text{F}^2=\sum_{i,j} A_{ij}^2= d$, so that $\gamma$ alone controls the intensity of the nonequilibrium drive. For convenience we assume that the spatial dimension $d$ is even, and choose the following form of the matrix $\bA$:
\begin{equation}\label{eq:A}
\gamma\bA \equiv \bigoplus_{\alpha=1}^{d/2} \bGamma
\end{equation}
with $\bm\Gamma \equiv \gamma\begin{bmatrix} 0 & -1 \\ 1 & 0 \end{bmatrix}$ a $2\times 2$ matrix. There is no loss of generality in choosing this form, since any antisymmetric matrix of even dimension $\gamma\bA$, with a spectrum $\{ \pm i\lambda_k \}_{k=1,\ldots,d/2}$ can be rewritten via an orthogonal transformation in a block form analogous to the one of Eq.~\eqref{eq:A}, with the $\gamma$'s replaced by $\gamma\lambda_k$ in each block. As will become clear in the course of the derivation, allowing for $\lambda_k\neq 1$ does not alter the general results discussed in the following sections (but it would require a minor alteration of some explicit formulas).

Due to the structure of $\bA$ it is convenient to decompose the vector $\bu_i = \bigoplus_{\alpha=1}^{d/2} \bu_{i,\alpha}$
\begin{equation}\label{eq:IO_infd_liquid_start}
    \zeta \dot{\mathbf{u}}_{i,\alpha}(t) = -(\mathds{1} + \bm\Gamma) \cdot \bnabla_{i.\alpha}U(t) + \bm{\xi}_{i,\alpha}(t) ,
\end{equation}
where for a given $d$-dimensional vector $\bx$ we have introduced its components $\bx_{\alpha}= (x_{2\alpha-1}, x_{2\alpha})^T\equiv (x_{\alpha}^{(1)}, x_{\alpha}^{(2)})^T$, a set of $d/2$ two-dimensional vectors containing the directions that affect each other non reciprocally via the transverse force. By construction it follows that $\bx = \bigoplus_{\alpha=1}^{d/2} \bx_{\alpha} $.

We first discuss the scaling of the various parameters with spatial dimension $d$ so that the resulting $d\to\infty$ dynamics retains interesting many-body features. First, we recall the specifics of the large $d$ limit of equilibrium dynamics. As always in mean-field, each particle interacts with typically $d$ neighbors. The notion of neighbor makes sense when the range $\ell$ of the pair interaction potential is finite, as assumed here. Then the typical number of neighbors per particle is $\rho V_d(\ell)$ where $V_d(\ell)=\frac{\pi^{d/2}}{\Gamma(d/2+1)}\ell^d$ is the volume of the interaction sphere. As $d\to\infty$ we want to maintain $\rho V_d(\ell)$ of order $d^1$. Then, to ensure that the energy is extensive in the number $dN$ of degrees of freedom, the pair potential $v(r)$ needs to admit the following scaling behavior,
\begin{equation}
    v(r) \equiv \overline{v}(h),\,h=d\left(\frac{r}{\ell}-1\right),
\end{equation}
where $\overline{v}(h)$ remains finite as $d\to\infty$. This implies that the forces scale as $v'(r) \sim d\overline{v}'(h) \sim d$. On the other hand, the displacement $\Delta r \equiv \lvert \mathbf{r}^{(a)}_0 + \mathbf{u}^{(a)} \rvert - r_0$ of a given particle is of order $d^{-1}$, as $h$ needs to be of order unity. It follows then from the expansion $\Delta r \approx \mathbf{\hat{r}} \cdot \mathbf{u} + \frac{\mathbf{u}^2}{2r_0}$ that $u^{(a)}_{\mu} \sim d^{-1}$. To keep all the terms of Eq.~\eqref{eq:IO_infd_liquid_start} of the same order in $d$, we need to scale the friction as $\zeta \sim d^2$. Finally, we need to discuss the main new element in Eq.~\eqref{eq:IO_infd_liquid_start}, \textit{i.e.} the transverse forces. With our choice \eqref{eq:A} of matrix $\bA$, in order for these forces to act in a nontrivial yet non singular way, we need $\gamma\sim d^0$. 

\section{One-particle process}\label{sec:1p}

We consider $N+1$ particles labeled by $i=0,\ldots,N$ and we now proceed to derive  in the infinite dimensional limit an exact equation for a tagged, appropriate degree of freedom pertaining to particle $0$. In an equilibrium setting, the authors of~\cite{liu2021dynamics} understood that the proper cavity variable could not be a full $d$-dimensional position vector (because the number of degrees of freedom of the cavity cannot be extensive in $d$). In our situation, however,  we identify the proper degree of freedom to be the vectors $\bu_{0,\alpha}=(u_{0,2\alpha-1}, u_{0,2\alpha})^T$, with $\alpha$ fixing the pair of space directions. 

With this identification, we start by writing a Liouville operator ${L}(t)$ governing the dynamics of the $N+1$ particles:
\begin{equation}\label{eq:L}
    \begin{split}
        L(t) &\equiv \sum_{i=0}^N \sum_{\mu=1}^{d/2} \frac{1}{\zeta}\Biggl[ -(\mathds{1}_2 + \bm\Gamma) \cdot \bnabla_{i,\mu} U (t) \\
        &+  \bm{\xi}_{i,\mu}(t) + T \bnabla_{i, \mu} \Biggr]^T \cdot \bnabla_{i, \mu}
    \end{split}
\end{equation}
for a given realization of the stochastic forces. Then one can write for a generic vector $\bx_{i,\mu}(t) = \mathcal{U}[L](t,0) \cdot \mathbf{x}_{i,\mu}$, where $\mathbf{x}_{i,\mu} \equiv \mathbf{x}_{i,\mu}(0)$ and\
\begin{equation}
     \mathcal{U}[L](t,0) \equiv \exp\left( \mathcal{T} \int_0^t d\tau L(\tau)\right).
\end{equation}
The operator $\mathcal{T}$ is the time ordering operator from left to right, so that $\p_t \mathcal{U}(t,0) = \mathcal{U}(t,0) L(t)$~\cite{j2007statistical}.

We associate to the Liouvillean operator an irreducible version $L^{\text{irr}}$:
\begin{equation}
    \begin{split}
        L^{\text{irr}}(t) &\equiv L(t) - \delta L(t) , \\
        \delta L(t) &\equiv \frac{1}{\zeta}\Biggl[ -(\mathds{1}_2 + \bm\Gamma) \cdot \bnabla_{0,\alpha} U(t) \\
        &+  \bm{\xi}_{0,\alpha}(t) + T \bnabla_{0, \alpha} \Biggr]^T \cdot \mathcal{P} \cdot \bnabla_{0, \alpha}
    \end{split}
\end{equation}
where the projection operator $\mathcal{P}$ is defined by
\begin{equation}
    \begin{split}
        \mathcal{P}\mathbf{x}(t) &\equiv \left\langle \frac{\int d\mathbf{r}^{\perp}_0 \prod_{i>0,\mu}d\mathbf{r}_{i,\mu} e^{-\beta U} \mathbf{x}(t) }{\int d\mathbf{r}^{\perp}_0 \prod_{i>0,\mu}  d\mathbf{r}_{i,\mu} e^{-\beta U} } \right\rangle_{\bm{\xi}_0,\ldots,\bm{\xi}_N} \\
        &\equiv \langle \bx(t) \rangle_0 ,
    \end{split}
\end{equation}
where $d\mathbf{r}^{\perp}_0 \equiv \prod_{\mu \neq \alpha} d\mathbf{r}_{0,\mu}$ indicates integration over all the component of particle $0$ with the exclusion of the pair of tagged directions $\alpha$. The average $\langle \cdots \rangle_{\bm{\xi}_0,\ldots,\bm{\xi}_N} \equiv \langle \ldots \rangle_0$ is an average over realizations of the noise of all the particle and coordinates, including the tagged one. In short, $\mathcal{P}$ averages over all the degrees of freedom except for the tagged ones.

We now express the force $\bnabla_{0,\alpha} U(t)$ using the irreducible Liouvillean. To do this, we exploit the identity
\begin{equation}\label{eq:identity_e^A+B}
    \begin{split}
 \mathcal{U}[L](t;0) &=\mathcal{U}[L^{\text{irr}}](t;0)\\
&+\int_0^t{\rm d}\tau \mathcal{U}[L](\tau;0)\delta L(\tau)\mathcal{U}[L^{irr}](t;\tau) ,
 \end{split}   
\end{equation}
to write
\begin{equation}\label{eq:F_irred}
    \begin{split}   
        & -\left( \mathds{1}_2 + \bm\Gamma \right) \cdot \bnabla_{0,\alpha} U(t)  \\ &=  -\left( \mathds{1}_2 + \bm\Gamma \right) \cdot \bnabla_{0,\alpha} U^\dagger(t) \\
        &- \int_0^t d\tau \mathcal{U}[L](\tau,0)\delta L(\tau)\mathcal{U}[L^{\text{irr}}](t,\tau)  \left( \mathds{1} + \bm\Gamma \right) \cdot \bnabla_{0,\alpha} U ,
    \end{split}
\end{equation}
where we defined $\bnabla_{0,\alpha} U^\dagger(t) \equiv \mathcal{U}[L^{\text{irr}}](t,0) \cdot  \bnabla_{0,\alpha} U$. We will show in the following that the first and second terms on the right hand side of Eq.~\eqref{eq:F_irred} are respectively a fluctuating force and an effective friction. In order to make this identification explicit, we simplify the term under  the integral of Eq.~\eqref{eq:F_irred} into
\begin{widetext}
\begin{equation}
    \begin{split}
         -\left\{\mathcal{U}[L](\tau,0)\delta L(\tau)\mathcal{U}[L^{\text{irr}}](t,\tau)  ( \mathds{1}_2 + \bm\Gamma ) \cdot \bnabla_{0,\alpha} U\right\}^{(a)} = -\beta\left\{(\mathds{1}_2 + \bm\Gamma) \cdot \left\langle \bnabla_{0,\alpha} U^\dagger(t-\tau) \otimes \bnabla_{0,\alpha} U^\dagger \right\rangle_0  \cdot \dot{\mathbf{u}}_{0,\alpha}(\tau) \right\}^{(a)} ,
    \end{split}
\end{equation}
\end{widetext}
where we recall that $u^{(a)}_\alpha$ is the $a$-th component of the two dimensional vector $\bu_\alpha$. Eq.~\eqref{eq:IO_infd_liquid_start} for the coordinate $\alpha$ of particle $0$ reads
\begin{equation}\label{eq:effeq_Fdagger}
\begin{split}
    &\zeta \dot{\mathbf{u}}_{0,\alpha}(t) = -(\mathds{1}_2 + \bm\Gamma)\cdot \bnabla_{0,\alpha} U^\dagger(t) \\ 
    &- \beta \int_0^t d\tau (\mathds{1}_2 + \bm\Gamma) \cdot   \left\langle \bnabla_{0,\alpha} U^\dagger(t-\tau) \otimes \bnabla_{0,\alpha} U^\dagger \right\rangle_0 \cdot \dot{\mathbf{u}}_{0,\alpha}(\tau) \\
    &+ \bm{\xi}_{0,\alpha}(t) .
\end{split}
\end{equation}
It is important to realize that so far Eq.~\eqref{eq:effeq_Fdagger} is exact. This equation expresses the decomposition of a force into the sum of an effective noise and of a friction term, as customarily obtained through the projection operator formalism~\cite{zwanzig1960ensemble}. However, we will show that in the infinite dimensional limit the memory kernel is further simplified by substituting $\bnabla_{0,\alpha} U^\dagger(t)$ with $\bnabla_{0,\alpha} \widetilde{U}(t)$, the force exerted by the fluid where coordinates $\alpha$ of particle $0$ have been frozen. To do so we first fully write the expression of $L^{\text{irr}}$, with $\mathcal{Q} \equiv 1 - \mathcal{P}$:
\begin{equation}\label{eq:L_irr_L0}
    \begin{aligned}
        &L^{\text{irr}}(t) = \frac{1}{\zeta}\left[-(\mathds{1}_2 + \bm\Gamma) \cdot \bnabla_{0,\alpha} U + \bm{\xi}_{0,\alpha}(t) + T\bnabla_{0,\alpha} \right]^{\text{T}} \cdot \mathcal{Q}\bnabla_{0,\alpha} \\ 
        &+ \sum_{\mu\neq\alpha}^{d/2}  \frac{1}{\zeta}\left[-(\mathds{1}_2 + \bm\Gamma) \cdot\bnabla_{0,\mu} U + \bm{\xi}_{0,\mu}(t) + T\bnabla_{0,\mu} \right]^{\text{T}} \cdot \bnabla_{0,\mu} \\
        &+  \sum_{i>0}^N\sum_{\mu=1}^{d/2}  \frac{1}{\zeta}\left[-(\mathds{1}_2 + \bm\Gamma) \cdot\bnabla_{i,\mu} U + \bm{\xi}_{i,\mu}(t) + T\bnabla_{i,\mu} \right]^{\text{T}} \cdot \bnabla_{i,\mu} \\
        &\equiv \frac{1}{\zeta}\left[-(\mathds{1}_2 + \bm\Gamma) \cdot \bnabla_{0,\alpha} U + \bm{\xi}_{0,\alpha}(t) + T\bnabla_{0,\alpha} \right]^{\text{T}} \cdot \mathcal{Q}\bnabla_{0,\alpha} \\ 
        &+ L^0(t) ,
    \end{aligned}
\end{equation}
from which we see that $\bnabla_{0,\alpha} \widetilde{U}(t) = \mathcal{U}[L^0](t,0)\bnabla_{0,\alpha} U$. By using the last line of Eqs.~(\ref{eq:identity_e^A+B}, \ref{eq:L_irr_L0}) we can express the force $\bnabla_{0,\alpha} U^\dagger(t)$ in terms of a Dyson series
\begin{equation}
    \begin{split}
        & \bnabla_{0,\alpha} U^\dagger(t) = \bnabla_{0,\alpha}\widetilde{U}(t) \\ 
        &+ \int_0^t d\tau \mathcal{U}[L^0](t,\tau) \frac{1}{\zeta}\Biggl[(\mathds{1}_2 + \bm\Gamma) \cdot \bnabla_{0,\alpha} U \\
        &+ \bm{\xi}_{0,\alpha}(\tau) + T\bnabla_{0,\alpha} \Biggr]^{\text{T}} \cdot \mathcal{Q}\bnabla_{0,\alpha} \mathcal{U}[L^0](\tau,0)\bnabla_{0,\alpha} U \\
        &+ \ldots
    \end{split}
\end{equation}
We will now argue that all the terms of this series excluding the first are sub-leading in the infinite dimensional limit, provided that this limit is taken for a finite time $t$. For example, the second term of the series can be rewritten as
\begin{equation}
    \begin{split}
        &\int_0^t d\tau \mathcal{U}[L^0](t,\tau) \frac{1}{\zeta}\biggl[(\mathds{1}_2 + \bm\Gamma) \cdot \bnabla_{0,\alpha} U \\
        &+ \bm{\xi}_{0,\alpha}(\tau) + T\bnabla_{0,\alpha} \biggr]^{\text{T}} \cdot \left[ \widetilde{\mathbf{k}}_\alpha(\tau) - \left\langle \widetilde{\mathbf{k}}_\alpha(\tau)\right\rangle_0\right],
    \end{split}
\end{equation}
where $\widetilde{k}^{(ab)}_\alpha(0) \equiv \sum_{j>0} \nabla_{\alpha}^{(a)} \nabla_{\alpha}^{(b)}v(r_{0j}^{(ab)}) \delta_{ab}$. The fluctuations of this term are of order $d^{3/2}$. Therefore for finite times the integral scales as $\frac{1}{d}d^{3/2} \sim d^{1/2}$, and is sub-leading compared to the first term of the series which is of order $d$. Terms of the series of order $n$ all contain additional powers of the friction coefficient $\zeta$, leading to the scaling $d^{3/2 - n}$. In conclusion, we can rewrite Eq. \eqref{eq:effeq_Fdagger} as 
\begin{equation}\label{eq:effeq_Ftilde}
\begin{split}
    & \zeta\dot{\mathbf{u}}_{0,\alpha}(t) = -(\mathds{1}_2 + \bm\Gamma)\cdot \bnabla_{0,\alpha} \widetilde{U}(t) \\
    &- \beta\int_0^t d\tau (\mathds{1}_2 + \bm\Gamma) \cdot   \left\langle \bnabla_{0,\alpha} \widetilde{U}(t-\tau) \otimes \bnabla_{0,\alpha} \widetilde{U} \right\rangle_0 \cdot \dot{\mathbf{u}}_{0,\alpha}(\tau) \\
    &+ \bm{\xi}_{0,\alpha}(t) .
\end{split} 
\end{equation}
We have thus obtained an effective equation for the degree of freedom $\mathbf{u}_{0,\alpha}(t)$, the coordinates $\alpha$ of the position for particle $0$. The force-force correlation can be rewritten as
\begin{equation}\label{eq:Mab}
  \begin{aligned}
    &\left\langle \nabla_{0,\alpha}^{(a)} \widetilde{U}(t)\nabla_{0,\alpha}^{(b)} \widetilde{U} \right\rangle_0
    = \left\langle \sum_{i,j \neq 0} \nabla_{0,\alpha}^{(a)} v(\widetilde{r}_{0i}(t))\nabla_{0,\alpha}^{(b)} v(\widetilde{r}_{0i})\right\rangle_0 \\
    &= \left\langle \sum_{i \neq 0}\nabla_{0,\alpha}^{(a)} v(\widetilde{r}_{0i}(t))\nabla_{0,\alpha}^{(b)} v(\widetilde{r}_{0i}) \right\rangle_0 \\
     &=  \frac{2}{d} \left\langle \sum_{\mu=1}^{d/2}\sum_{i \neq 0}\nabla_{0,\mu}^{(a)} v(\widetilde{r}_{0i}(t))\nabla_{0,\mu}^{(b)} v(\widetilde{r}_{0i}) \right\rangle_0 \\
    &= \frac{2}{dN} \left\langle \sum_{\mu=1}^{d/2}\sum_{i \neq 0} \nabla_{0,\mu}^{(a)} v({r}_{0i})(t) \nabla_{0,\mu}^{(b)} v({r}_{0i}) \right\rangle \\
    &\equiv M^{(ab)}(t).
  \end{aligned}
\end{equation}
The indices $a$ and $b$ identify the cavity degrees for the two coupled tagged space directions, as defined after Eq.~\eqref{eq:IO_infd_liquid_start}. In deriving Eq.~\eqref{eq:Mab} we have used, in order:
\begin{itemize}
  \item the fact that forces between particle $0$ and other two distinct particles $i$ and $j$ are uncorrelated in the large dimensional limit;
  \item the equivalent roles played by the $d/2$ blocks;
  \item the fact that in the large dimensional limit the average $\langle ... \rangle_{0,\alpha}$ over the dynamics with the frozen direction $\alpha$ of particle $0$ can be replaced by an average over the dynamics of the full, unfrozen system, and that such an average can be carried out for all the $N$ particles of the system. 
\end{itemize}
Moreover, we observe that in the high dimensional limit the force $\bnabla_{0,\alpha} \widetilde{U}(t)$ has a zero average,  and its statistics are entirely determined by its second moment. The dynamical equation obtained for particle $0$ can be generalized to any particle in the liquid, since they are all identical. The resulting equation reads
\begin{equation}
  \begin{aligned}
    &\zeta\mathbf{\dot u}_{i, \alpha} = -\beta \int_0^t d\tau ( \mathds{1} + \bm{\Gamma} ) \cdot \mathbf{M}(t-\tau) \mathbf{\dot u}_{i,\alpha}(\tau) + \bm{\Xi}_{i,\alpha}(t) \label{eq:sp} \\
    &\langle \bm{\Xi}_{i, \alpha}(t) \rangle = 0 \\
    &\left\langle \bm{\Xi}_{i,\alpha}(t) \otimes \bm{\Xi}_{j,\beta}(t') \right\rangle = \delta_{ij}\delta_{\alpha\beta}\bigg[ 2T\zeta\mathds{1}_d\delta(t-t') \\
    &+ ( \mathds{1}_2 + \bm{\Gamma} ) \cdot \mathbf{M}(t - t') \cdot ( \mathds{1}_2 + \bm{\Gamma^T} ) \bigg].
  \end{aligned}
\end{equation}
The memory kernel $\bM(t)$ encoding pairwise force-force correlations is self-consistently determined by the dynamics of two interacting particles. The latter can be obtained from the microscopic dynamics in the same way as in Eq.~\eqref{eq:sp}. However, we first observe that the memory kernel at the initial time is completely determined by the equilibrium properties of the system. Therefore, it has the same expression as its equilibrium counterpart, and in particular it is diagonal:
\begin{equation}\label{eq:M0}
\bM(0) = \mathds{1}_2 \frac{\rho}{d}\int d\br_0 g(\br_0) \lvert v'(r_0) \rvert^2,
\end{equation}
with $g(\br) = e^{-\beta v(r)}$ the radial distribution function in the infinite-dimensional limit. The physical simplifying idea that we use to proceed is that a diagonal kernel at initial times implies, in the large dimensional limit, a diagonal kernel at successive times.  In the following Section, we will derive the equations of motion for the two particle process under the self consistent assumption of a diagonal kernel $\bM = \mathds{1}_2M(t)$. In this case the dynamics for the full displacement of particle $i$ in $d$ dimensions reads
\begin{equation}\label{eq:oneparticle}
    \begin{split}
        \zeta \bu_i(t) &= -\beta\int_0^t d\tau M(t-\tau) (\mathds{1}_d + \gamma \bA)\cdot \dot\bu_i(\tau) + \bXi(t) \\
        \left\langle \bXi_i(t) \otimes \bXi_j(t)\right\rangle &= \mathds{1}_d\delta_{ij}\left[ 2T\zeta\delta(t-t') + (1+\gamma^2)M(t-t')\right] \\
        M(t) &= \frac{\rho}{d}\int d\br_0 g(\br_0)\left\langle \brhat(t)v'(r(t))\right\rangle_{\br_0}\cdot \brhat_0 v'(r_0) ,
    \end{split}
\end{equation}
where the definition of the kernel $M(t)$ comes from an extension by continuity of Eq. \eqref{eq:M0}. 

\section{Two-particle process}\label{sec:2p}

\subsection{General formulation}

We consider two particles, which we label  $0$ and $1$, and we focus on the dynamics of their separation $\br(t) \equiv \bR_0(t) - \bR_1(t)$. A derivation analogous to the one performed for the one-particle process, along the lines of \cite{liu2021dynamics}, gives:
\begin{equation}\label{eq:twobody}
    \begin{split}
        \frac{\zeta}{2}\dot \br &= -(\mathds{1}_d + \gamma \bA)\brhat v'(r(t)) \\
        &- \frac{\beta}{2}\int_0^t d\tau M(t-\tau)(\mathds{1}_d + \gamma \bA)\dot\br(\tau) + \sqrt{2}\bXi(t) \\
        \left\langle \bXi(t) \otimes \bXi(t')\right\rangle &= \mathds{1}_d\left[ 2T\zeta\delta(t-t') + (1+\gamma^2)M(t-t')\right].
    \end{split}
\end{equation}
Equation~\eqref{eq:twobody} can intuitively be derived from the one-particle process of Eq.~\eqref{eq:oneparticle}, singling the force between particles $0$ and $1$ out of the expression of $M$. This is a legitimate operation which yields a negligible correction in the large dimensional limit.

From Eq.~\eqref{eq:twobody} using It\=o calculus one can obtain equations of motion for the dynamics of the distance $r=||\br||$ and the unit vector $\brhat$ along $\br$:
\begin{equation}\label{eq:radial_angular}
    \begin{split}
        &\frac{\zeta}{2}\dot r(t) = \frac{T(d-1)}{r(t)} - \frac{\beta}{2}\int_0^t d\tau M(t-\tau)\brhat(t)\cdot(\mathds{1}_d + \gamma \bA)\dot\br(\tau) \\
        &+ \sqrt{2}\brhat(t)\cdot\bXi(t) \\
        &\frac{\zeta}{2}\mathbf{\dot{\hat{r}}}(t) = -\frac{T(d-1)}{r(t)^2}\brhat(t) - \gamma \bA \brhat(t) v'(r(t)) \\
        &-\frac{\beta}{2r(t)}\int_0^t d\tau M(t-\tau)\cdot \mPiort{t}\left(\mathds{1}_d + \gamma \bA\right)\dot\br(\tau) \\
        &+ \sqrt{2}\mPiort{t}\cdot \bXi(t) ,
    \end{split}
\end{equation}
with $\mPiort{t} \equiv \mathds{1}_d -\brhat(t)\otimes\brhat(t)$ the operator projecting along a direction orthogonal to $\brhat(t)$. At $t=0$ the dynamics of $\brhat$ reads
\begin{equation}
    \begin{split}
        \frac{\zeta}{2}\mathbf{\dot{\hat{r}}}(0) &= -\frac{T(d-1)}{r(0)^2}\brhat(0) - \gamma \bA \brhat(0) v'(r(0)) \\
        &+ \sqrt{2}\mPiort{t}\cdot \bXi(0).
    \end{split}
\end{equation}
The left hand side of this equations scales as $d^{3/2}$, while the right hand side contains terms of order at most $d$. We therefore conclude that  the orientation of the vector $\br$ does not change to leading order in $d$ with respect to its initial orientation. We can thus assume that $\brhat(t) = \brhat(0)$ at all times. Then the evolution of $\brhat(t)$ in  Eq.~\eqref{eq:radial_angular} becomes
\begin{equation}
    \begin{split}
        \frac{\zeta}{2}\mathbf{\dot{\hat{r}}}(t) &= -\frac{T(d-1)}{r(t)^2}\brhat(0) - \gamma \bA \brhat(0) v'(r(t)) \\
        &-\frac{\gamma \beta}{2r(t)}\int_0^t d\tau M(t-\tau)\bA\dot\br(\tau) \\
        &+ \sqrt{2}\mPiort{t} \cdot \bXi(t).
    \end{split}
\end{equation}
The scaling for the different terms reads as above, and the new term introduced, which is of order $d^2\times d^{-1}=d$, remains subleading. This proves self-consistently that, as was the case in equilibrium, the orientation of the inter particle separation is constant throughout the dynamics, which was not an {\it a priori} obvious property. The equation for the interparticle distance reads:
\begin{equation}\label{eq:twobodyradial}
    \begin{split}
        \frac{\zeta}{2}\dot r(t) &= -v'(r(t)) - \frac{T(d-1)}{r(t)} \\
        &- \frac{\beta}{2}\int_0^t d\tau M(t-\tau)r(\tau) + \sqrt{2}\Xi(t) \\
        \left\langle \Xi(t)\Xi(t')\right\rangle &= 2T\zeta\delta(t-t') + (1+\gamma^2)M(t-t') \\
        M(t) &= \frac{\rho \Omega_d}{d}\int \dd r_0 g(r_0)\langle v'(r(t))\rangle_{r_0} v'(r_0) .
    \end{split}
\end{equation}
Note that the two-particle process is still an out of equilibrium one, since the noise and the friction terms do not respect the fluctuation-dissipation theorem. One expects therefore a different value of the memory kernel $M(t)$ compared to the $\gamma=0$ case. The analytical determination of $M(t)$ at all times is an open question, even in the equilibrium case, where progress has been made either by numerical integration or a low density expansion~\cite{manacorda2020numerical}.  We turn to the latter approach, in order to produce an expression of $M(t)$ which will be useful in the following sections, where we probe the efficiency of the transverse forces.  

\subsection{Low density expansion of the memory kernel}

The main idea is to expand the memory kernel in the form of a perturbative series
\begin{equation}
    M(t) = \sum_{n=1}^{\infty} M^{(n)}(t) ,
\end{equation}
where $M^{(n)} \sim O(\rho^{n})$ can be self-consistently determined from the two particle process given by Eq.~\eqref{eq:twobodyradial}, evaluated up to order $O(\rho^{n-1})$. Therefore, the lowest order $M^{(1)}$ is determined by the process
\begin{equation}\label{eq:twobodyradial_rho0}
    \begin{split}
        \frac{\zeta}{2}\dot r^{(0)}(t) &= \frac{T(d-1)}{r^{(0)}(t)} - v'(r^{(0)}(t)) + \sqrt{2}\Xi^{(0)}(t) \\
        \left\langle \Xi^{(0)}(t)\Xi^{(0)}(t')\right\rangle &= 2T\zeta\delta(t-t'),
    \end{split}
\end{equation}
which is the same as the one obtained at equilibrium~\cite{manacorda2020numerical}. Low densities suppress the action of transverse forces at the level of the two-body process. In particular, one can consider the case of a linear potential $v(r) = \epsilon(\frac{r}{\ell}-1)$, for which $M^{(1)}(t)$ was determined in Ref.~\cite{manacorda2020numerical}. We give here the expression of its time integral $\widehat{M^{(1)}}\equiv \int_0^{+\infty} \dd t M^{(1)}(t)$, which will be useful later. It reads
\begin{equation}\label{eq:M_lowphi}
    \beta\widehat{M^{(1)}} = \frac{\widehat{\phi}}{2}\frac{\beta^2(2+\beta)}{(1+\beta)^3} ,
\end{equation}
with $\widehat\phi \equiv \rho V_d \frac{\ell^d}{d}$ the rescaled packing fraction, and with $V_d$ the volume of a sphere of unit radius in $d$ dimensions. Below, we will address the issues of the long time limit of $M$ and of its asymptotic behavior as $\gamma$ goes to infinity.

\section{Dynamical arrest}

\label{sec:dynarr}

\subsection{The  transition occurs at the same location as in equilibrium}
The glass transition in mean field fluids can be found by looking at the long time behavior of $M(t)$. Following the existing literature~\cite{gotze2009complex, parisi2020theory} we split the kernel $M$ into a decaying part and an asymptotic plateau value at $t$ becomes large: $M(t)=M_f(t) + M_p$, with $M_f(t\to \infty)=0$ and $M_p\geq 0$ a constant. A glass transition occurs when the density or the temperature of the system are such that $M_p> 0$. In this Section we will show that this happens at the same parameters as for equilibrium dynamics.

The plateau value $M_p$ is given by:
\begin{equation}\label{eq:Mp}
M_p =\lim_{t\to\infty}M(t) = \frac{\rho}{d}\int d\br_0 \left\langle v'(r)\right\rangle_{\text{ss}} v'(r_0), 
\end{equation}
where $\left\langle\ldots\right\rangle_{\text{ss}}$ is an average over the steady state value of the dynamics for the distance between two particles, given by Eq. \eqref{eq:twobody}. Such a steady state distribution depends on the value of $M_p$, thus yielding a self-consistent relation that can be exploited to determine the plateau value. 

Using the aforementioned decomposition for $M(t)$ the equation of motion for $\br$ can be rewritten as a nonequilibrium dynamics for a particle in $d$ dimensions inside an effective potential $w(\br)$, under the action of a constant Gaussian drift $\bXi_p$ (induced by the long time dynamics of $M(t)$ and fluctuating Gaussian noises $\bXi_f$, $\bxi$) with eventually decaying correlations:
\begin{equation}\label{eq:effr}\begin{split}
    \frac{\zeta}{2}\dot\br &= (\mathds{1}_d + \gamma \bA)\cdot\bigg[-\bnabla w(\br) \\
    &-\frac{\beta}{2}\int_0^t d\tau M_f(t-\tau)\dot\br(\tau) + \bXi_f(t)\bigg] \\
    &+ \sqrt{T}\bxi(t) ,
\end{split}\end{equation}
with
\begin{equation}
    w(\br) =v(\lvert\br\rvert) + \frac{\beta}{4}M_p\left[ \br - \br_0 \right]^2 + \bXi_p\cdot\br,
\end{equation}
the correlations of the constant random drive $\bXi_p$ and the time dependent noises $\bXi_f(t)$, $\bxi(t)$ being respectively $\left\langle \bxi(t)\otimes\bxi(t')\right\rangle = \mathds{1}_d T\zeta \delta(t-t')$, $\left\langle \bXi_f(t) \otimes \bXi_f(t')\right\rangle = \mathds{1}_d\frac{1}{2}M_f(t-t')$, $\left\langle \bXi_p \otimes \bXi_p \right\rangle = \mathds{1}_d\frac{1}{2}M_p$.

We briefly review the equilibrium case ($\gamma=0$) for which we have  
\begin{equation}\label{eq:effreq}\begin{split}
    \frac{\zeta}{2}\dot\br &= -\bnabla w(\br) -\frac{\beta}{2}\int_0^t d\tau M_f(t-\tau)\dot\br(\tau) \\
    &+ \bXi_f(t) + \sqrt{T}\bxi(t).
\end{split}\end{equation}
This is an equilibrium dynamics with memory under an external potential $w(\br)$. The equilibrium distribution is the Boltzmann one:
\begin{equation}
p_{\text{\tiny{B}}}(\br\lvert M_p,\bXi_p, \br_0) \equiv \frac{e^{-\beta w(\br)}}{\int d\br e^{-\beta w(\br)}}.
\end{equation}
We note that the steady state distribution depends on the plateau value of the memory kernel, and it is conditioned on the realizations of the field $\bXi_f$ and on the initial condition $\br_0$. The steady state value of the force appearing in the self-consistent equation Eq.~\eqref{eq:Mp} is
\begin{equation}\begin{split}
\left\langle v'(r)\right\rangle_{\text{eq},\br_0} &= \frac{1}{\left(\pi M_p\right)^{d/2}}\int d\bXi_p e^{-\frac{\bXi_p^2}{M_p}} \\
&\times\int d\br p_{\text{\tiny{B}}}(\br,\lvert M_p,\bXi_p, \br_0) v'(\lvert \br \rvert).
\end{split}\end{equation}
Substitution in Eq.~\eqref{eq:Mp} yields the desired self-consistent relation. We refer the reader to Refs.~\cite{maimbourg2016solution, parisi2020theory} for a detailed discussion of its solution. Here it is sufficient to use the fact that Eq.~\eqref{eq:Mp} admits a nonzero value for $M_p$ below a critical temperature (or above a critical density $\rho_d$) $T_d$, meaning that the system is no longer ergodic below $T_d$ (above $\rho_d$). 

After having reviewed the equilibrium limit, we now return to the $\gamma\neq0$ case of interest. The steady state distribution of the process given by Eq.~\eqref{eq:effr} is the same as the equilibrium one. This means that the self-consistent relation given by Eq.~\eqref{eq:Mp} yields a nonzero value of $M_p$ for the same critical parameter as in the equilibrium dynamics, thus proving the statement that opened this Section. This reflects the fact that the transverse force dynamics is constructed to preserve the Boltzmann distribution in the steady state. Any ergodicity breaking that takes its root in the thermodynamics, as the one observed in mean field fluids, will be observed also in the presence of transverse forces at the same point as in equilibrium. However, even if the glass transition point is unchanged, it is interesting to observe how the dynamics conspires to produce this result, and how it differs from its equilibrium counterpart.

\subsection{A sanity check}

The spirit of the dynamical mean-field theory is to integrate out in an exact fashion an extensive number of degrees of freedom. This integration step transforms the Markovian dynamics into one with memory. From an analysis of the dynamics, it is thus not easy to see that the steady-state distribution for the non-Markovian process is the Boltzmann one. We know however that this must be the case, because the integration can be done at the static level using replicas \cite{parisi2020theory}, and we know that in equilibrium, the analysis of the dynamics~\cite{maimbourg2016solution} confirms the results of the statics. This subsection may thus seem a somewhat superfluous sanity check, but it is in principle needed. We want to prove directly the invariance of the steady state distribution of Eq.~\eqref{eq:effr}. We express the memory kernel $M_f(t)$ as a sum of exponentials:
\begin{equation}
\frac{1}{2}M_f(t) = \sum_k c_k e^{-t/\tau_k} ,
\end{equation}
where the $c_k$'s and the $\tau_k$'s are appropriately distributed~\cite{POTTIER20112863}. Using this decomposition, we can rewrite the non-Markovian equation of motion for $\br$ as a Markovian one, at the cost of introducing an extra set of degrees of freedom, $\by_k$, coupled to $\br$:
\begin{equation}\label{eq:ry}
    \begin{split}
        &\frac{\zeta}{2}\dot\br(t) = -(\mathds{1}_d+\gamma\bA)\cdot \bnabla w(\br) \\
        &+ \sum_k \sqrt{c_k\beta}(\mathds{1}_d+\gamma\bA)\cdot \left[ \by_k(t) - \sqrt{c_k \beta}\left[\br(t) -\br_0\right] \right] \\
        &+ \sqrt{T} \bxi , \\
        &\dot\by_k = -\frac{1}{\tau_k} \left[ \by_k - \sqrt{c_k\beta}\left[ \br(t) - \br_0\right] \right]+ \sqrt{\frac{2T}{\tau_k}}\beeta_k \\
        &\left\langle \beeta_i(t) \otimes \beeta_j(t')\right\rangle = \mathds{1}_d\delta_{ij}\delta(t-t') \\
        &\left\langle \bxi_i(t) \otimes \bxi_j(t')\right\rangle = \mathds{1}_d\delta(t-t') .
    \end{split}
\end{equation}
We choose $\by_k(0)$ to be independent random variables drawn from a Gaussian distribution of variance $T$  for all $k$. Upon averaging over the Markovian  evolution and the initial condition of the $\by_k$ variables, Eq.~\eqref{eq:ry} is then identical to Eq.~\eqref{eq:effr}. The dynamics of $\by_k$ reads:
\begin{equation}\begin{split}
    \by_k(t) &= \by_k(0)e^{-t/\tau_k} + \int_0^t d\tau e^{-\frac{t-\tau}{\tau_k}} \\
    &\times\left[ \frac{\sqrt{c_k\beta}}{\tau_k} \left[\br(\tau) -\br_0\right] + \sqrt{\frac{2T}{\tau_k}}\beeta_k(\tau) \right],
\end{split}\end{equation}
which, after an integration by parts, becomes
\begin{equation}
    \begin{split}
        \by_k(t) &=  \by_k(0)e^{-t/\tau_k} + \sqrt{c_k\beta}\left[ \br(t) - \br_0 \right] \\
        &- \sqrt{c_k\beta}\int_0^t d\tau e^{-\frac{t-\tau}{\tau_k}}\dot\br(\tau) + \sqrt{\frac{2T}{\tau_k}}\int_0^t d\tau e^{-\frac{t-\tau}{\tau_k}} \beeta_k(\tau). 
    \end{split}
\end{equation}
Substitution into the equation for $\br$ yields
\begin{equation}
    \begin{split}
        \frac{\zeta}{2}\dot\br(t) &= -(\mathds{1}_d + \gamma\bA)\bnabla w(\br) \\
        &- \beta \int_0^t d\tau \sum_k c_k e^{-\frac{t-\tau}{\tau_k}}(\mathds{1}_d + \gamma\bA)\cdot \dot\br(\tau) \\
        &+ (\mathds{1}_d + \gamma\bA)\cdot \bnu(t) + \sqrt{T}\bxi(t) \\
        \bnu(t) &\equiv \sum_k \sqrt{c_k\beta}\by_k(0) e^{-t/\tau_k} + \sqrt{\frac{2c_k}{\tau_k}}\int_0^t d\tau e^{-\frac{t-\tau}{\tau_k}} \beeta_k(\tau) .
    \end{split}
\end{equation}
The second term on the right hand side is $\frac{\beta}{2}\int_0^td\tau M_f(t-\tau)(\mathds{1}_d + \gamma\bA)\cdot \dot\br(\tau)$. The contribution $\bnu(t)$ is a Gaussian noise whose correlations are given by
\begin{equation}
    \left\langle\bnu(t)\otimes\bnu(t')\right\rangle = \mathds{1}_d \sum_k c_k e^{\frac{\lvert t-t' \rvert}{\tau_k}} = \frac{1}{2}\mathds{1}_d M_f(\lvert t-t'\rvert),
\end{equation}
and therefore $\bnu(t)=\bXi_f(t)$. This concludes the proof of equivalence between Eq.~\eqref{eq:effr} and Eq.~\eqref{eq:ry}.   

To obtain the steady state distribution of the joint process for $\br$ and $\by$, we first rewrite the equation in the following form:
\begin{equation}
    \begin{split}
        \frac{\zeta}{2}\dot\br(t) &= -(\mathds{1}_d + \gamma\bA) \left[ \bnabla w(\br) + \bnabla \widetilde w(\br,\left\{\by_k\right\}) \right] + \sqrt{T}\bxi(t) \\
        \by_k(t) &= -\frac{1}{\tau_k}\bpartial_k \widetilde w(\br,\left\{\by_k\right\}) + \sqrt{\frac{2T}{\tau_k}}\beeta(t) ,
    \end{split}
\end{equation}
where $\widetilde w(\br,\left\{\by_k\right\}) = \sum_k \frac{1}{2}\left( \by_k - \sqrt{c_k\beta}\br\right)^2$ and $\bpartial_k \equiv \frac{\partial}{\partial \by_k}$. This is an overdamped Langevin dynamics under the action of an external potential $w+\widetilde w$ and transverse forces acting on the $\br$ variables. Therefore, it admits the steady state distribution 
\begin{equation}
p_{\text{ss}}(\br,\left\{\by_k\right\}\lvert M_p,\bXi_p, \br_0) \propto e^{-\beta\left(w +  w(\br,\left\{\by_k\right\}\right)} .
\end{equation}
Upon integrating out the auxiliary variables $\by_k$, we obtain
\begin{equation}
p_{\text{ss}}(\br\lvert M_p,\bXi_p, \br_0) = \frac{e^{-\beta w(\br)}}{\int d\br e^{-\beta w(\br)} } = p_{\text{\tiny{B}}}(\br\lvert M_p,\bXi_p, \br_0).
\end{equation}
This result implies a self-consistent equation for $M_p$ identical to the one holding in equilibrium, and therefore an identical glass transition temperature. This maybe does not come as a surprise given that, in equilibrium, at least in infinite dimensions, the glass transition point can also be found by resorting to thermodynamic methods~\cite{parisi2020theory}, and the transverse forces are designed to preserve the thermodynamics.

\section{Ergodic phase with strong nonequilibrium drive}

\label{sec:gammalarge}

The calculation of the previous Section has demonstrated that the behavior of the plateau value of the force-force correlation for transverse forces is identical to the one for equilibrium dynamics. However, in the presence of transverse forces the two-particle process at any finite time is different from its equilibrium counterpart, and we therefore expect $M(t)$ to be affected by the nonequilibrium forces in the ergodic region. The explicit determination of the memory kernel for arbitrary temperature and density is out of reach, and already in equilibrium the numerical integration of the equation of motion and the self-consistent relation is a formidable task~\cite{manacorda2020numerical}. Here, we will address the scaling of the memory kernel $M(t)$ in the limit $\gamma\to\infty$.

The starting point is the dynamics of the interparticle distance:
\begin{equation}
    \begin{split}
        \frac{\zeta}{2}\dot r(t) &= \frac{T(d-1)}{r(t)} -v'(r(t)) \\ 
        &-\frac{\beta}{2}\int_0^t d\tau M(t-\tau)\dot r(\tau) + \Xi(t) \\
        \left\langle \Xi(t)\Xi(t')\right\rangle &= T\zeta\delta(t-t') + \frac{1}{2}(1+\gamma^2)M(t-t').
    \end{split}
\end{equation}
We recall that the memory kernel $M(t)$ is determined from the two-body dynamics itself, see Eq.~\eqref{eq:twobodyradial}. 

We now rescale time, $\overline t \equiv \gamma t$, and we define the new functions $\overline f(\overline t) \equiv f\left(\frac{\overline t}{\gamma}\right)$, obtaining
\begin{equation}
    \begin{split}
        \gamma \frac{\zeta}{2}\dot{\overline r}\left(\overline t\right) &= \frac{T(d-1)}{\overline r\left(\overline t\right)} -v'\left(\overline r\left(\overline t\right)\right) \\
        &-\frac{\beta}{2}\int_0^{\frac{\overline t}{\gamma}} d\tau \overline M\left(\overline t-\tau\right)\dot{\overline r}(\tau) + \overline \Xi\left(\overline t\right) \\
        \left\langle \overline \Xi\left(\overline t\right)\overline \Xi\left(\overline t'\right)\right\rangle &= \gamma T\zeta\delta\left(\overline t-\overline t'\right) + \frac{1}{2}(1+\gamma^2)\overline M\left(\overline t-\overline t'\right) ,
    \end{split}
\end{equation}
with $\overline M\left(\overline t\right) = \frac{\rho}{d}\int d\br_0 g\left(\br_0 \right)\left\langle v'\left(\overline r\left(\overline t\right)\right)\right\rangle_{\br_0} v'(r_0)$. If we now send $\gamma\to \infty$ and keep only the leading terms, we obtain
\begin{equation}\label{eq:twobodystrongdrive}
    \begin{split}
        \frac{\zeta}{2}\dot{\overline r}\left(\overline t\right) &= \overline\Xi\left(\overline t\right) \\
        \left\langle \overline \Xi\left(\overline t\right)\overline \Xi\left(\overline t'\right)\right\rangle &= \frac{1}{2}\overline M\left(\overline t - \overline t'\right) .
    \end{split}
\end{equation}
Note that in the rescaled units this equation is no longer dependent on $\gamma$, and as a consequence $\overline M\left(\overline t\right)$ cannot depend on $\gamma$. Using this result, we can determine the scaling of the zero frequency mode of the memory kernel, $\widehat M(0)$:
\begin{equation}
    \begin{split}
        \widehat M(0) &= \int_0^{+\infty} M(t)dt \\
        &= \frac{1}{\gamma}\int_0^{+\infty} M\left(\frac{\overline t}{\gamma}\right)d\overline t \\
        &= \frac{1}{\gamma}\int_0^{+\infty} \overline M\left(\overline t\right) d\overline t\sim \frac{1}{\gamma}.
    \end{split}
\end{equation}
We have thus found the large $\gamma$ behavior of the zero-frequency mode of the memory kernel.

\section{Mean-squared displacement and diffusion constant}

\label{sec:msd}

In this Section, we explore how the transverse forces dynamics influence the diffusivity of the particles. We are interested in the mean-squared displacement
\begin{equation}
    \Delta(t) \equiv \frac{1}{N}\sum_i \left\langle \left[ \bR_i(t) - \bR_i(0) \right]^2 \right\rangle = \left\langle \bu_0(t)^2 \right\rangle . 
\end{equation}
In the long time limit we expect diffusive behavior
\begin{equation}
    \lim_{t\to\infty} \Delta(t) = 2dD_\parallel(T,\gamma)t.
\end{equation}
Our aim is to obtain an expression for $D_\parallel(T,\gamma)$. The starting point of our calculation is the one-particle dynamics (we omit the particle index since they are all equivalent)
\begin{equation}\label{eq:onep_2}
    \zeta \dot\bu(t) = -\int_0^t d\tau M(t-\tau)\left(\mathds{1}_d + \gamma \bA\right)\dot\bu(\tau) + \bXi(t) ,
\end{equation}
with the noise correlations $\left\langle \bXi(t) \otimes \bXi(t') \right\rangle = \mathds{1}_d\left[2T\zeta\delta(t-t') + (1+\gamma^2)M(t-t')\right]$. Introducing the Laplace transform $\widehat f(s) \equiv \int_0^{+\infty}dt e^{-st}f(t)$, we can write the mean-squared displacement using the Bromwich inversion integral
\begin{equation}\label{eq:deltabromwich}
    \Delta(t) = -\frac{1}{4\pi^2}\int_{z-i\infty}^{z+i\infty}\int_{z-i\infty}^{z+i\infty} ds ds' e^{st + s't} \left\langle \mathbf{\widehat{u}}(s) \cdot \mathbf{\widehat{u}}(s') \right \rangle ,
\end{equation}
with $z$ greater than the real part of all the singularities of the integrand. Applying the Laplace transform to both sides of Eq.~\eqref{eq:onep_2} and noting that $\bA\bA^T = \mathds{1}_d$ we get 
\begin{equation}\label{eq:us}
    \begin{split}
        \bu(s) &= \frac{\left(\zeta + \beta\widehat M(s)\right)\mathds{1}_d - \gamma\beta\widehat M(s) \bA}{\left(\zeta + \beta\widehat M(s)\right)^2 + \left(\gamma\beta\widehat M(s)\right)^2}\cdot \frac{\boldsymbol{\widehat{\Xi}}(s)}{s} \\
        &\equiv \mathbf{\widehat{K}}(s)\cdot \frac{\boldsymbol{\widehat{\Xi}}(s)}{s}.
    \end{split}
\end{equation}
We also need to know the noise correlations in Laplace space, which read
\begin{equation}
    \begin{split}
        \left\langle \mathbf{\widehat{\Xi}}(s) \otimes \mathbf{\widehat{\Xi}^{\text{T}}}(s') \right\rangle &= \mathds{1}_d \frac{2T\zeta + (1 + \gamma^2)\left( \widehat M(s) + \widehat M(s')\right)}{s + s'} \\
        &\equiv \mathds{1}_d \frac{C(s,s')}{s+s'}.
    \end{split}
\end{equation}
Equation~\eqref{eq:deltabromwich} now reads
\begin{equation}\begin{split}
    \Delta(t) &= -\frac{1}{4\pi^2}\int_{z-i\infty}^{z+i\infty}\int_{z-i\infty}^{z+i\infty} ds ds' \frac{e^{st + s't}}{ss'(s+s')}\\ 
    &\times C(s,s')\Tr[\mathbf{\widehat K}(s)\cdot\mathbf{\widehat K^T}(s')] .
\end{split}\end{equation}
Since we are interested in the large time limit, we make the change of variables $w \equiv st$, $w' \equiv s't$:
\begin{equation}\begin{split}
     \Delta(t) &= -\frac{t}{4\pi^2}\int_{zt-i\infty}^{zt+i\infty}\int_{zt-i\infty}^{zt+i\infty} dw dw' \frac{e^{w + w'}}{ww'(w+w')} \\
     &\times C\left(\frac{w}{t},\frac{w'}{t}\right)\Tr[\mathbf{\widehat K}\left(\frac{w}{t}\right)\cdot\mathbf{\widehat K^T}\left(\frac{w'}{t}\right)] .
\end{split}\end{equation}
The diffusion constant is obtained sending $t\to+\infty$:
\begin{equation}
    \begin{split}
            D_\parallel(\gamma,T) &= \frac{\Tr[\mathbf{\widehat K}(0)\cdot\mathbf{\widehat K^T}(0)]C(0,0)}{2d} \\ 
            &\times \lim_{t\to+\infty }\left( \frac{1}{4\pi^2}\right)\int_{zt-i\infty}^{zt+i\infty}\int_{zt-i\infty}^{zt+i\infty} dw dw' \\
            &\times \frac{e^{w + w'}}{ww'(w+w')} .
    \end{split}
\end{equation}
Carrying out the computation explicitly yields
\begin{equation}\label{eq:D}
     D_\parallel(\gamma,T) =  T\frac{\zeta + (1+\gamma^2)\beta\widehat{M}(0)}{\left( \zeta + \beta\widehat M(0) \right)^2 + \left(\gamma\beta\widehat M(0)\right)^2} .
\end{equation}
This is the central result of this Section. For $\gamma=0$ we obtain the known equilibrium expression of the diffusion constant:
\begin{equation}
  D_\parallel(0,T) = \frac{T}{\zeta + \beta \widehat M(0)} .
\end{equation}
The nonequilibrium diffusion constant is always larger than the equilibrium one, $D_\parallel(\gamma,T) \geq D_\parallel(0,T)$, as long as $\widehat M(0)$ is equal to or smaller than its equilibrium counterpart. For large nonequilibrium drives, the zero-frequency mode of the memory kernel scales as $\gamma^{-1}$ (see Sec. \ref{sec:gammalarge} above), and the diffusion constant grows linearly with $\gamma$: 
\begin{equation}
  D_\parallel(T,\gamma \rightarrow \infty) \sim c(T)\gamma, 
\end{equation}
with $c(T)$ a temperature-dependent coefficient.

Concerning the temperature behavior of $D_\parallel$, two limits are of interest. The first one is the infinite temperature limit. In this case $\widehat M(0)=0$ and one obtains
\begin{equation}
    D_\parallel(T\to\infty,\gamma) = \frac{T}{\zeta} = D_\parallel(T\to\infty,0).
\end{equation}
Since thermal fluctuations dominate over the interactions when $T\to\infty$, the transverse forces fail to accelerate the dynamics, and the nonequilibrium nature of the process is washed out by the thermal noise. The second limit of interest is for $T\to T_d$, the glass transition temperature. Here $\widehat M(0)$ diverges and the diffusion constant goes to $0$, signaling dynamical arrest. 

To conclude this Section, we compare explicitly $D_\parallel(\gamma,T)$ and its equilibrium counterpart in the low density regime for a linear potential, where $\widehat{M}(0)$ takes the form given in Eq.~\eqref{eq:M_lowphi}. The results are shown in Fig.~\ref{fig:Drel_lowphi}(a). The efficiency of transverse forces, defined as the ratio$\frac{D_\parallel(\gamma,T)}{D_\parallel(\gamma,0)}$ changes non monotonically with the temperature for the highest values of $\gamma$. This feature survives also in finite dimensions, as recently evidenced in extensive numerical simulations~\cite{ghimenti2023sampling}.

Our investigation of the diffusion constant suggests the following picture: upon reducing the temperature from the $T\to \infty$ regime dominated by thermal noise, transverse forces emerge from thermal fluctuations and accelerate the dynamics. However, as memory effects become stronger and the glass transition is approached, the enhancement of the diffusion is reduced. We are thus led to the question of what is the effect of transverse forces in the low temperature regime. In the next Section we show that they mostly give rise to odd transport coefficients rather than providing a stronger dynamical speedup.  

\section{Emergent odd transport}

\subsection{Odd diffusivity}\label{sec:odddiff}

In this subsection we prove, and quantify, the presence of odd diffusion in the infinite dimensional fluid driven by transverse foces. Odd diffusion manifests itself in the form of a nonzero off-diagonal, anti-symmetric part of the diffusion tensor. Physically, this detects the presence of chiral, swirling motion, and the presence of fluxes perpendicular to concentration gradients arising in the system. At the microscopic level, a Green-Kubo relation, derived in~\cite{hargus2021odd} identifies the odd diffusion as the time integral of the anti-symmetric part of the velocity-velocity autocorrelation tensor. Following this approach, we have
\begin{equation}
    \begin{split}
        \bD(\gamma,T) &\equiv \frac{1}{N}\sum_{i=1}^N \int_0^{+\infty} \dd t \langle \dot\bu_i(t)\otimes \dot\bu_i(0)\rangle \\
        &= D_\parallel(\gamma,T)\mathds{1}_d + D_\perp(\gamma,T) \bA .
    \end{split}
\end{equation}
The conventional longitudinal diffusion constant $D_\parallel(\gamma,T)$ given by Eq.~\eqref{eq:D} appears in the diagonal entries of the diffusion tensor. The anti-symmetric contribution is proportional to the odd diffusion constant $D_\perp(T,\gamma)$. It is defined as: 
\begin{equation}\label{D_perp equiv}
  \begin{aligned}
    D_\perp &\equiv \frac{1}{d N}\sum_{i=1}^N \int_0^{\infty}\left\langle \dot\bu_{i}(t) \cdot \bA\dot\bu_{i}(0)\right\rangle \\
    &= \frac{1}{\zeta N d}\sum_{i=1}^N\lim_{z\to0}\left\langle s\mathbf{\widehat{u}}(s) \cdot\bA\bXi_{i}(0) \right\rangle\\
    &=  \frac{1}{\zeta d}\lim_{s\to0}\left\langle s\mathbf{\widehat{u}}(s)\cdot\bA\bXi(0) \right\rangle.
  \end{aligned}
\end{equation}
In the second equality we have used a representation in terms of the Laplace transform of the displacement $\bu_{i,\alpha}$, and in the third equality the fact that all particles are equivalent. Using Eq.~\eqref{eq:us} and the fact that:
\begin{equation}
    \left\langle \boldsymbol{\widehat{\Xi}}(s)\otimes \bXi(0)\right\rangle = \mathds{1}_d T\left[ \zeta + (1+\gamma^2)\beta\widehat M(s)\right] 
\end{equation}
we obtain
\begin{equation}
    D_\perp = -\frac{T}{\zeta d} \Tr[\mathbf{\widehat{K}}(0) \bA]\left[ \zeta + (1+\gamma^2)\beta\widehat M(0)\right] ,
\end{equation}
with $\mathbf{\widehat{K}}(0)$ defined in Eq.~\eqref{eq:us}. An explicit computation of the trace yields
\begin{equation}
D_\perp = -\frac{T\gamma}{\zeta}\frac{\beta \widehat M(0) + (1+\gamma^2)\left(\beta\widehat M(0)\right)^2}{\left( \zeta + \beta\widehat M(0) \right)^2 + \left(\gamma\beta\widehat M(0)\right)^2},
\end{equation}
thus proving the presence of a finite odd diffusivity in the system whenever $\gamma\neq0$. Upon approaching the glass transition, the odd diffusivity converges to the nonzero value $\gamma\frac{T}{\zeta}$: even when ergodicity is broken and the system is confined in a long-lived metastable state, a form of odd transport persists, analogous to the swirling motion of a particle trapped in an harmonic well in the presence of transverse forces~\cite{ghimenti2023sampling}. 

As we shall see now, odd diffusion is not the only emergent odd transport coefficient.

\subsection{Odd mobility}\label{sec:oddmob}

Odd mobility \cite{poggioli2023odd} is a transport coefficient closely related to the odd diffusivity. In this subsection we show that its behavior upon approaching the glass transition is distinctly different from the odd diffusivity as we find that odd mobility vanishes in the nonergodic phase.

Odd mobility describes the transverse motion of a tracer upon applying a constant force. Specifically, we consider the case where a constant force $\bFext$ is applied at $t=0$ on particle $0$, which thus assumes the role of a tracer. The system thus evolves under the action of the operator $\Lext(t)$, defined as
\begin{equation}
    \Lext(t) = L(t) + \frac{1}{\zeta}\sum_{\beta=1}^{d/2} \bFext_{\beta} \cdot \bnabla_{0,\beta},
\end{equation}
with $L(t)$ the evolution operator of the unperturbed system, displayed in Eq.~\eqref{eq:L}. Following a linear response formalism, we have, to first order in $\bFext$,
\begin{equation}\label{eq:evolution_linear_response}
    \begin{split}
    \mathcal{U}[\Lext](t,0) &\approx \mathcal{U}[L](t,0) \\
    &+ \frac{1}{\zeta}\sum_{\beta=1}^{d/2}\int_0^{t} \dd\tau \mathcal{U}[L](t,\tau) \bFext_{\beta} \cdot \bnabla_{0,\beta} \mathcal{U}[L](0,\tau). 
    \end{split}
\end{equation}
To obtain the equation of motion of the perturbed tracer in the linear response, we have to compute $\mathcal{U}[\Lext](t,0)\bF_{0,\alpha}$. However, we observe that $\mathcal{U}[L](t,0)\bF_{0,\alpha}$ is translationally invariant. This implies that only the first term on the right hand side of Eq.~\eqref{eq:evolution_linear_response} contributes to the evolution of $\bF_{0,\alpha}$, and therefore
\begin{equation}
    \mathcal{U}[\Lext](t,0)\bF_{0,\alpha}(t) = \mathcal{U}[L](t,0)\bF_{0,\alpha} + O((\bFext)^2).
\end{equation}
The equation of motion of the perturbed tracer in the infinite dimensional limit thus becomes
\begin{equation}\label{eq:tracer}
    \begin{split}
        \zeta \dot \bu_i(t) &= \bFext -\int_0^t d\tau M(t-\tau) (\mathds{1}_d + \gamma \bA)\cdot \dot\bu_i(\tau) \\
        &+ \bXi(t) ,
    \end{split}
\end{equation}
with the noise $\bXi(t)$ and the memory kernel $M(t)$ corresponding to the one of the unperturbed dynamics in Eq.~\eqref{eq:oneparticle}. The mobility of the tracer is defined via the relation
\begin{equation}
    \lim_{t\to+\infty}\langle \dot\bu_0(t)\rangle = \bmu\bFext.
\end{equation}
Applying a Laplace transform to Eq.~\eqref{eq:tracer}, taking the 0-frequency limit and keeping the leading diverging terms we obtain the following expression for $\bmu$:
\begin{equation}\label{eq:mobility}
    \begin{split}
    \bmu &= \frac{\left[\zeta+\beta \widehat{M}(0)\right]\mathds{1} - \gamma\beta\widehat{M}(0)\bA}{\left[\zeta + \beta\widehat{M}(0)\right]^2 + \left(\gamma\beta\widehat{M}(0)\right)^2} \\
    &\equiv \mu_\parallel \mathds{1} + \mu_\perp \bA.
    \end{split}
\end{equation}
For $\gamma=0$ we fall back to the equilibrium case, $\bmu=\frac{1}{\zeta + \beta \widehat{M}(0)}\mathds{1}$ and the Einstein relation is satisfied, $T \bmu =\bD$. When $\gamma \neq 0$, the Einstein relation breaks down. Note in passing that this violation takes a compact form for the longitudinal component of the diffusivity tensor, namely ${\mathbf{D}}_\parallel=[(\mathds{1}+\gamma \mathbf{A})\bmu]_\parallel$. The mobility is composed of a longitudinal term, $\mu_\parallel$, and of an odd component, $\mu_\perp$. The ratio $\frac{D_\parallel}{T\mu_\parallel}=1+\gamma^2\frac{\beta\widehat{M}(0)}{1+\beta \widehat{M}(0)}$ is greater than one, which hints at a more efficient exploration of configurations than in equilibrium, but this effective temperature~\cite{cugliandolo1997energy} is not the one that drives dynamical arrest. 

\begin{figure}
    \includegraphics[width=\columnwidth]{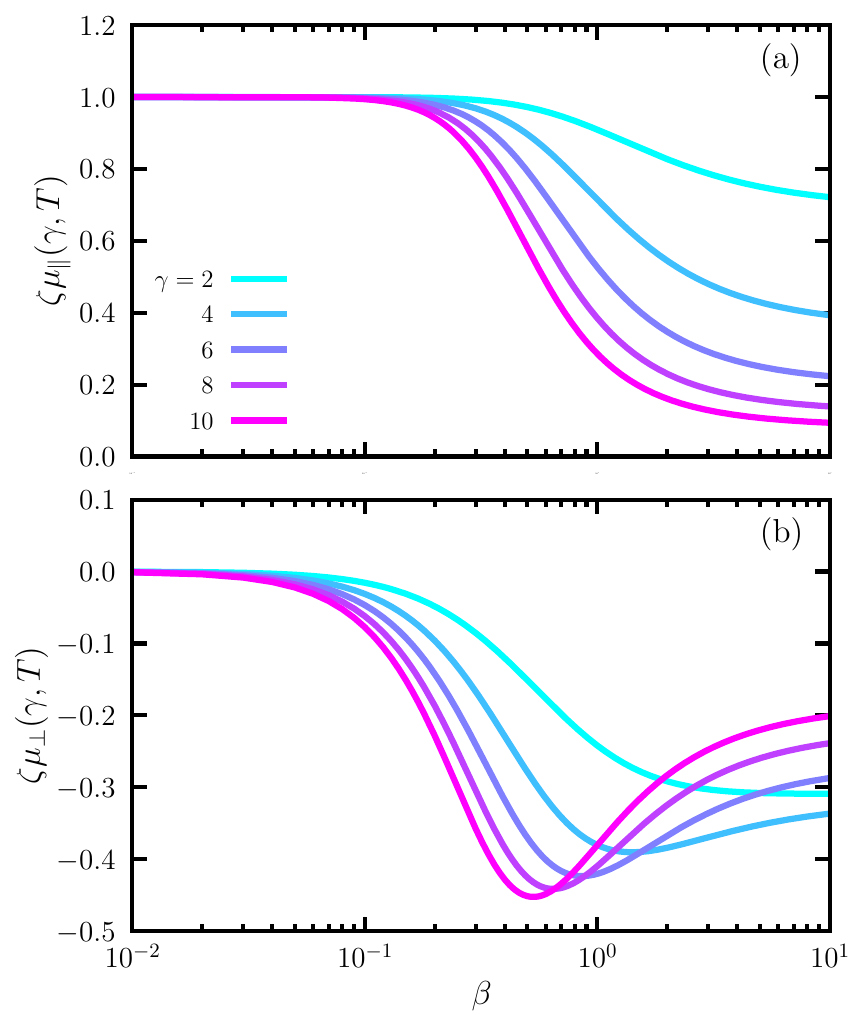}
    \caption{(a) Longitudinal mobility $\mu_\parallel$ for different values of the strength $\gamma$ of the non equilibrium drive, as a function of the inverse temperature $\beta\equiv T^{-1}$. (b) Odd mobility in the presence of transverse forces. In  both panels, the memory kernel used is the one obtained via a low density expansion for the case of a linear potential. Its expression is given in Eq.~\eqref{eq:M_lowphi}.}
    \label{fig:mobility}
\end{figure}

A plot of longitudinal and transverse mobilities as a function of the inverse temperature is shown in Fig.~\ref{fig:mobility}, where the low density approximation of the memory kernel $\widehat{M}(0)\approx \widehat{M}^{(1)}$ [see Eq.~\eqref{eq:M_lowphi}] was used. The longitudinal mobility decreases with temperature from its free particle high-$T$ value, while the modulus of the odd mobility has a non-monotonic behavior, with a maximum located in the same region where the odd diffusion steeply rises, and where the efficiency of transverse forces starts decreasing, which was depicted in Fig.~\ref{fig:Drel_lowphi}. This simultaneous occurrence of a similar behavior in all these quantities supports the physical picture of transverse forces operating at their best in the mildly interacting regime.

We also see from Eq.~\eqref{eq:mobility} that both the longitudinal and transverse mobilities vanish at the glass transition $T_d$. Physically, the above result implies the absence of a long-time systematic displacement in the non-ergodic phase both along the direction of the external force and in the direction transverse to it. 

\subsection{Odd viscosity}

\label{sec:oddvisc}

This subsection is devoted to the computation of the odd viscosity. Odd viscosity appears as an anti-symmetric component of the viscous tensor. Upon application of an external shear stress, a system with odd viscosity develops a flow in the plane orthogonal to the one along which the shear stress is applied. A Green-Kubo approach, developed in Ref.~\cite{epstein2020time}, relates the odd viscosity of a nonequilibrium system to the time integral of stress-stress correlation functions. Within this framework, using linear response theory, the viscous tensor in the hydrodynamic limit reads
\begin{equation}
    \eta_{abcd} = \frac{\beta}{V}\int_0^{+\infty} dt \left\langle \sigma_{ab}(t)\sigma^{\text{IK}}_{cd}(0)\right\rangle .
\end{equation}
The average of the integrand is meant both with respect to the initial condition and with respect to the realizations of the noise. The notation $\sigma^{\text{IK}}_{ab}$ refers to the Irving-Kirkwood stress tensor for overdamped dynamics:
\begin{equation}
    \sigma^{\text{IK}}_{ab} = -\frac{1}{2}\sum_{i \neq j} \frac{r_{ij,a}(t)r_{ij,b}(t)}{r_{ij}(t)}v'(r_{ij}(t)) ,
\end{equation}
and $\sigma_{ab}$ is the stress tensor in the presence of transverse forces
\begin{equation}
    \sigma_{ab} = (\mathds{1}_d + \gamma\bA)_{bc}\sigma^{\text{IK}}_{ac}.
\end{equation}
The odd viscosity is finally defined as 
\begin{equation}
    \eta_\perp \equiv \frac{1}{2}\left(\eta_{xyxx} - \eta_{xxxy}\right) ,
\end{equation}
where $x=2a-1$ and $y=2a$ with $a=1,\ldots,d/2$. To be explicit, we focus on the $x=0$, $y=1$ case. The odd viscosity then becomes
\begin{equation}
    \begin{split}
        \eta_\perp &= \frac{\beta}{2V}\int_0^{+\infty}dt \left\langle \sigma^{\text{IK}}_{xy}(t)\sigma^{\text{IK}}_{xx}(0)\right\rangle - \left\langle \sigma^{\text{IK}}_{xx}(t)\sigma^{\text{IK}}_{xy}(0)\right\rangle 
        \\ &+ \gamma \left\langle \sigma^{\text{IK}}_{xx}(t)\sigma^{\text{IK}}_{xx}(0)\right\rangle + \gamma \left\langle \sigma^{\text{IK}}_{xy}(t)\sigma^{\text{IK}}_{xy}(0)\right\rangle .
    \end{split}
\end{equation}
The first two terms inside the integrand vanish because they are odd upon rotation in the $xy$ plane. The latter terms are
\begin{widetext}
\begin{equation}
    \begin{split}
        &\left\langle \sigma^{\text{IK}}_{xx}(t)\sigma^{\text{IK}}_{xx}(0)\right\rangle + \left\langle \sigma^{\text{IK}}_{xy}(t)\sigma^{\text{IK}}_{xy}(0)\right\rangle \\ 
        &=\frac{1}{4} \left\langle \sum_{i\neq j, k\neq l}\frac{r_{ij,x}(t)r_{ij,x}(t)r_{kl,x}(0)r_{kl,x}(0) + r_{ij,x}(t)r_{ij,y}(t)r_{kl,x}(0)r_{kl,y}(0)}{r_{ij,x}(t)r_{kl,x}(0)}v'(r_{ij}(t))v'(r_{kl}(0))\right\rangle \\
        &= \frac{1}{4} \left\langle \sum_{i\neq j}\frac{r_{ij,x}(t)r_{ij,x}(t)r_{ij,x}(0)r_{ij,x}(0) + r_{ij,x}(t)r_{ij,y}(t)r_{ij,x}(0)r_{ij,y}(0)}{r_{ij,x}(t)r_{ij,x}(0)}v'(r_{ij}(t))v'(r_{ij}(0))\right\rangle \\
        &\sim \frac{1}{2d^2} \left\langle \sum_{i]\neq j}\frac{\lvert \br_{ij}\cdot\br_0\rvert^2}{r_{ij}(t)r_{ij}(0)}v'(r(t))v'(r_{ij}(0)) \right\rangle \\
        &= \frac{\rho^2 V}{2d^2}\int d\br_0 g(\br_0) \left \langle \frac{\lvert \br(t)\cdot\br_0\rvert^2}{r_{ij}(t)r_{ij}(0)}v'(r(t))v'(r_{ij}(0)) \right\rangle_{\text{dyn}} \\
        &= \frac{\rho V \ell^2}{2d} M(t),
    \end{split}
\end{equation}
\end{widetext}
where $\langle \ldots \rangle_{\text{dyn}}$ is an average over the realization of the noise in the two-body dynamics. In the last passage we have used the fact that the direction of the interparticle separation is constant throughout the dynamics and that to leading order in $d$ we have $r(t)\sim \ell$, with $\ell$ the characteristic interaction length of the potential. The odd viscosity is therefore
\begin{equation}\label{eq:oddviscosity}
    \eta_\perp = \frac{\gamma\beta\rho_0 \ell^2}{4d}\widehat{M}(0).
\end{equation}
As the glass transition is approached, the odd viscosity diverges. The physical interpretation of this phenomenon is that infinitely long-lived memory develops also at the level of stress fluctuations with respect to transverse perturbations, a consequence of the chiral interaction produced by the nonequilibrium drive. Below $T_d$, small external shear stresses applied to the system generate neither longitudinal nor transverse flows, consistently with the picture of a dynamically arrested glass, and both viscosities are formally infinite.

\section{Conclusion and outlook}

Working in the limit of a large number of space dimensions allows us to formulate a self-consistent dynamical mean-field theory for the nonequilibrium problem of a dense fluid submitted to both transverse and conservative forces. Unlike previous attempts with nonequilibrium  systems~\cite{agoritsas2019out1,agoritsas2019out2}, where the only possible progress is of numerical nature (in the spirit of \cite{manacorda2020numerical}), here the stationary state is known from the start. This knowledge is instrumental in being able to derive some of the dynamical properties, especially regarding the formal expression of the memory kernel $M$ that appears throughout. This remarkable feature has thus been very helpful in pushing our understanding of the mechanisms by which transverse forces are able to accelerate the dynamics. For instance, the reason why transverse forces struggle to accelerate the dynamics in the glassy regime can be understood in terms of the picture of particles falling, and swirling, into a local potential well. Our results show that increasing the amplitude of the transverse forces will formally lead to increased acceleration, but beyond a given threshold, this acceleration is physically equivalent to a rescaling of the mobility (thus effectively changing the time units without affecting the underlying physics). At more moderate values, however, transverse forces open up new dynamical pathways that qualitatively change the nature of the dynamics. Our coupled mean-field equations, for which the stationary state is known, are perhaps the simplest instance of nonequilibrium dynamics for which numerical integration is feasible, in the footsteps of \cite{manacorda2020numerical}. This is certainly a direction we wish to explore.

This work can also be seen as a first step towards exploring somewhat more involved acceleration methods such as lifting (this will change the exact definition of the cavity variable). It is not clear how the existence of odd transport in our transverse forces will translate in the presence of auxiliary lifting degrees of freedom, and what the interplay with acceleration will look like. 

In our work, the transverse force which is applied has the simplest possible structure. There exist other forms of transverse driving forces that can be imposed and finding the one that optimizes acceleration~\cite{lelievre2013optimal} is a key open question in many-body systems. While we expect our results to generalize readily to any antisymmetric matrix $\bA$, other, possibly more physics-informed choices for the nonequilibrium drive are possible. As for the present case, we can hope that the framework of dynamical mean-field theory will allow for progress in this interesting direction. 

\begin{acknowledgments}
LB, FG and FvW acknowledge the financial support of the ANR THEMA AAPG2020 grant. GS acknowledges the support of NSF Grant No.~CHE 2154241.
\end{acknowledgments}

\bibliography{final_biblio}

\begin{thebibliography}{49}%
\makeatletter
\providecommand \@ifxundefined [1]{%
 \@ifx{#1\undefined}
}%
\providecommand \@ifnum [1]{%
 \ifnum #1\expandafter \@firstoftwo
 \else \expandafter \@secondoftwo
 \fi
}%
\providecommand \@ifx [1]{%
 \ifx #1\expandafter \@firstoftwo
 \else \expandafter \@secondoftwo
 \fi
}%
\providecommand \natexlab [1]{#1}%
\providecommand \enquote  [1]{``#1''}%
\providecommand \bibnamefont  [1]{#1}%
\providecommand \bibfnamefont [1]{#1}%
\providecommand \citenamefont [1]{#1}%
\providecommand \href@noop [0]{\@secondoftwo}%
\providecommand \href [0]{\begingroup \@sanitize@url \@href}%
\providecommand \@href[1]{\@@startlink{#1}\@@href}%
\providecommand \@@href[1]{\endgroup#1\@@endlink}%
\providecommand \@sanitize@url [0]{\catcode `\\12\catcode `\$12\catcode
  `\&12\catcode `\#12\catcode `\^12\catcode `\_12\catcode `\%12\relax}%
\providecommand \@@startlink[1]{}%
\providecommand \@@endlink[0]{}%
\providecommand \url  [0]{\begingroup\@sanitize@url \@url }%
\providecommand \@url [1]{\endgroup\@href {#1}{\urlprefix }}%
\providecommand \urlprefix  [0]{URL }%
\providecommand \Eprint [0]{\href }%
\providecommand \doibase [0]{https://doi.org/}%
\providecommand \selectlanguage [0]{\@gobble}%
\providecommand \bibinfo  [0]{\@secondoftwo}%
\providecommand \bibfield  [0]{\@secondoftwo}%
\providecommand \translation [1]{[#1]}%
\providecommand \BibitemOpen [0]{}%
\providecommand \bibitemStop [0]{}%
\providecommand \bibitemNoStop [0]{.\EOS\space}%
\providecommand \EOS [0]{\spacefactor3000\relax}%
\providecommand \BibitemShut  [1]{\csname bibitem#1\endcsname}%
\let\auto@bib@innerbib\@empty
\bibitem [{\citenamefont {Ediger}\ \emph {et~al.}(1996)\citenamefont {Ediger},
  \citenamefont {Angell},\ and\ \citenamefont {Nagel}}]{ediger1996supercooled}%
  \BibitemOpen
  \bibfield  {author} {\bibinfo {author} {\bibfnamefont {M.~D.}\ \bibnamefont
  {Ediger}}, \bibinfo {author} {\bibfnamefont {C.~A.}\ \bibnamefont {Angell}},\
  and\ \bibinfo {author} {\bibfnamefont {S.~R.}\ \bibnamefont {Nagel}},\
  }\bibfield  {title} {\bibinfo {title} {Supercooled liquids and glasses},\
  }\href@noop {} {\bibfield  {journal} {\bibinfo  {journal} {The journal of
  physical chemistry}\ }\textbf {\bibinfo {volume} {100}},\ \bibinfo {pages}
  {13200} (\bibinfo {year} {1996})}\BibitemShut {NoStop}%
\bibitem [{\citenamefont {Berthier}\ and\ \citenamefont
  {Biroli}(2011)}]{berthier2011theoretical}%
  \BibitemOpen
  \bibfield  {author} {\bibinfo {author} {\bibfnamefont {L.}~\bibnamefont
  {Berthier}}\ and\ \bibinfo {author} {\bibfnamefont {G.}~\bibnamefont
  {Biroli}},\ }\bibfield  {title} {\bibinfo {title} {Theoretical perspective on
  the glass transition and amorphous materials},\ }\href@noop {} {\bibfield
  {journal} {\bibinfo  {journal} {Reviews of modern physics}\ }\textbf
  {\bibinfo {volume} {83}},\ \bibinfo {pages} {587} (\bibinfo {year}
  {2011})}\BibitemShut {NoStop}%
\bibitem [{\citenamefont {Berthier}\ and\ \citenamefont
  {Reichman}(2023)}]{berthier2023modern}%
  \BibitemOpen
  \bibfield  {author} {\bibinfo {author} {\bibfnamefont {L.}~\bibnamefont
  {Berthier}}\ and\ \bibinfo {author} {\bibfnamefont {D.~R.}\ \bibnamefont
  {Reichman}},\ }\bibfield  {title} {\bibinfo {title} {Modern computational
  studies of the glass transition},\ }\href@noop {} {\bibfield  {journal}
  {\bibinfo  {journal} {Nature Reviews Physics}\ }\textbf {\bibinfo {volume}
  {5}},\ \bibinfo {pages} {1} (\bibinfo {year} {2023})}\BibitemShut {NoStop}%
\bibitem [{\citenamefont {Barrat}\ and\ \citenamefont
  {Berthier}(2023)}]{barrat2023computer}%
  \BibitemOpen
  \bibfield  {author} {\bibinfo {author} {\bibfnamefont {J.-L.}\ \bibnamefont
  {Barrat}}\ and\ \bibinfo {author} {\bibfnamefont {L.}~\bibnamefont
  {Berthier}},\ }\bibfield  {title} {\bibinfo {title} {Computer simulations of
  the glass transition and glassy materials},\ }\href@noop {} {\bibfield
  {journal} {\bibinfo  {journal} {Comptes Rendus. Physique}\ }\textbf {\bibinfo
  {volume} {24}},\ \bibinfo {pages} {1} (\bibinfo {year} {2023})}\BibitemShut
  {NoStop}%
\bibitem [{\citenamefont {Grigera}\ and\ \citenamefont
  {Parisi}(2001)}]{grigera2001fast}%
  \BibitemOpen
  \bibfield  {author} {\bibinfo {author} {\bibfnamefont {T.~S.}\ \bibnamefont
  {Grigera}}\ and\ \bibinfo {author} {\bibfnamefont {G.}~\bibnamefont
  {Parisi}},\ }\bibfield  {title} {\bibinfo {title} {Fast monte carlo algorithm
  for supercooled soft spheres},\ }\href@noop {} {\bibfield  {journal}
  {\bibinfo  {journal} {Physical Review E}\ }\textbf {\bibinfo {volume} {63}},\
  \bibinfo {pages} {045102} (\bibinfo {year} {2001})}\BibitemShut {NoStop}%
\bibitem [{\citenamefont {Berthier}\ \emph {et~al.}(2016)\citenamefont
  {Berthier}, \citenamefont {Coslovich}, \citenamefont {Ninarello},\ and\
  \citenamefont {Ozawa}}]{berthier2016equilibrium}%
  \BibitemOpen
  \bibfield  {author} {\bibinfo {author} {\bibfnamefont {L.}~\bibnamefont
  {Berthier}}, \bibinfo {author} {\bibfnamefont {D.}~\bibnamefont {Coslovich}},
  \bibinfo {author} {\bibfnamefont {A.}~\bibnamefont {Ninarello}},\ and\
  \bibinfo {author} {\bibfnamefont {M.}~\bibnamefont {Ozawa}},\ }\bibfield
  {title} {\bibinfo {title} {Equilibrium sampling of hard spheres up to the
  jamming density and beyond},\ }\href@noop {} {\bibfield  {journal} {\bibinfo
  {journal} {Physical review letters}\ }\textbf {\bibinfo {volume} {116}},\
  \bibinfo {pages} {238002} (\bibinfo {year} {2016})}\BibitemShut {NoStop}%
\bibitem [{\citenamefont {Ninarello}\ \emph {et~al.}(2017)\citenamefont
  {Ninarello}, \citenamefont {Berthier},\ and\ \citenamefont
  {Coslovich}}]{ninarello2017models}%
  \BibitemOpen
  \bibfield  {author} {\bibinfo {author} {\bibfnamefont {A.}~\bibnamefont
  {Ninarello}}, \bibinfo {author} {\bibfnamefont {L.}~\bibnamefont
  {Berthier}},\ and\ \bibinfo {author} {\bibfnamefont {D.}~\bibnamefont
  {Coslovich}},\ }\bibfield  {title} {\bibinfo {title} {Models and algorithms
  for the next generation of glass transition studies},\ }\href@noop {}
  {\bibfield  {journal} {\bibinfo  {journal} {Physical Review X}\ }\textbf
  {\bibinfo {volume} {7}},\ \bibinfo {pages} {021039} (\bibinfo {year}
  {2017})}\BibitemShut {NoStop}%
\bibitem [{\citenamefont {Berthier}\ \emph {et~al.}(2019)\citenamefont
  {Berthier}, \citenamefont {Flenner}, \citenamefont {Fullerton}, \citenamefont
  {Scalliet},\ and\ \citenamefont {Singh}}]{berthier2019efficient}%
  \BibitemOpen
  \bibfield  {author} {\bibinfo {author} {\bibfnamefont {L.}~\bibnamefont
  {Berthier}}, \bibinfo {author} {\bibfnamefont {E.}~\bibnamefont {Flenner}},
  \bibinfo {author} {\bibfnamefont {C.~J.}\ \bibnamefont {Fullerton}}, \bibinfo
  {author} {\bibfnamefont {C.}~\bibnamefont {Scalliet}},\ and\ \bibinfo
  {author} {\bibfnamefont {M.}~\bibnamefont {Singh}},\ }\bibfield  {title}
  {\bibinfo {title} {Efficient swap algorithms for molecular dynamics
  simulations of equilibrium supercooled liquids},\ }\href@noop {} {\bibfield
  {journal} {\bibinfo  {journal} {Journal of Statistical Mechanics: Theory and
  Experiment}\ }\textbf {\bibinfo {volume} {2019}},\ \bibinfo {pages} {064004}
  (\bibinfo {year} {2019})}\BibitemShut {NoStop}%
\bibitem [{\citenamefont {Krauth}(2006)}]{krauth2006statistical}%
  \BibitemOpen
  \bibfield  {author} {\bibinfo {author} {\bibfnamefont {W.}~\bibnamefont
  {Krauth}},\ }\href@noop {} {\emph {\bibinfo {title} {Statistical mechanics:
  algorithms and computations}}},\ Vol.~\bibinfo {volume} {13}\ (\bibinfo
  {publisher} {OUP Oxford},\ \bibinfo {year} {2006})\BibitemShut {NoStop}%
\bibitem [{\citenamefont {Hwang}\ \emph {et~al.}(1993)\citenamefont {Hwang},
  \citenamefont {Hwang-Ma},\ and\ \citenamefont
  {Sheu}}]{hwang1993accelerating}%
  \BibitemOpen
  \bibfield  {author} {\bibinfo {author} {\bibfnamefont {C.-R.}\ \bibnamefont
  {Hwang}}, \bibinfo {author} {\bibfnamefont {S.-Y.}\ \bibnamefont
  {Hwang-Ma}},\ and\ \bibinfo {author} {\bibfnamefont {S.-J.}\ \bibnamefont
  {Sheu}},\ }\bibfield  {title} {\bibinfo {title} {Accelerating gaussian
  diffusions},\ }\href@noop {} {\bibfield  {journal} {\bibinfo  {journal} {The
  Annals of Applied Probability}\ ,\ \bibinfo {pages} {897}} (\bibinfo {year}
  {1993})}\BibitemShut {NoStop}%
\bibitem [{\citenamefont {Hwang}\ \emph {et~al.}(2005)\citenamefont {Hwang},
  \citenamefont {Hwang-Ma},\ and\ \citenamefont
  {Sheu}}]{hwang2005accelerating}%
  \BibitemOpen
  \bibfield  {author} {\bibinfo {author} {\bibfnamefont {C.-R.}\ \bibnamefont
  {Hwang}}, \bibinfo {author} {\bibfnamefont {S.-Y.}\ \bibnamefont
  {Hwang-Ma}},\ and\ \bibinfo {author} {\bibfnamefont {S.-J.}\ \bibnamefont
  {Sheu}},\ }\bibfield  {title} {\bibinfo {title} {Accelerating diffusions},\
  }\href@noop {} {\bibfield  {journal} {\bibinfo  {journal} {The Annals of
  Applied Probability}\ }\textbf {\bibinfo {volume} {15}},\ \bibinfo {pages}
  {1433} (\bibinfo {year} {2005})}\BibitemShut {NoStop}%
\bibitem [{\citenamefont {Ichiki}\ and\ \citenamefont
  {Ohzeki}(2013)}]{ichiki2013violation}%
  \BibitemOpen
  \bibfield  {author} {\bibinfo {author} {\bibfnamefont {A.}~\bibnamefont
  {Ichiki}}\ and\ \bibinfo {author} {\bibfnamefont {M.}~\bibnamefont
  {Ohzeki}},\ }\bibfield  {title} {\bibinfo {title} {Violation of detailed
  balance accelerates relaxation},\ }\href@noop {} {\bibfield  {journal}
  {\bibinfo  {journal} {Physical Review E}\ }\textbf {\bibinfo {volume} {88}},\
  \bibinfo {pages} {020101} (\bibinfo {year} {2013})}\BibitemShut {NoStop}%
\bibitem [{\citenamefont {Vucelja}(2016)}]{vucelja2016lifting}%
  \BibitemOpen
  \bibfield  {author} {\bibinfo {author} {\bibfnamefont {M.}~\bibnamefont
  {Vucelja}},\ }\bibfield  {title} {\bibinfo {title} {Lifting—a nonreversible
  markov chain monte carlo algorithm},\ }\href@noop {} {\bibfield  {journal}
  {\bibinfo  {journal} {American Journal of Physics}\ }\textbf {\bibinfo
  {volume} {84}},\ \bibinfo {pages} {958} (\bibinfo {year} {2016})}\BibitemShut
  {NoStop}%
\bibitem [{\citenamefont {Krauth}(2021)}]{krauth2021event}%
  \BibitemOpen
  \bibfield  {author} {\bibinfo {author} {\bibfnamefont {W.}~\bibnamefont
  {Krauth}},\ }\bibfield  {title} {\bibinfo {title} {Event-chain monte carlo:
  foundations, applications, and prospects},\ }\href@noop {} {\bibfield
  {journal} {\bibinfo  {journal} {Frontiers in Physics}\ }\textbf {\bibinfo
  {volume} {9}},\ \bibinfo {pages} {663457} (\bibinfo {year}
  {2021})}\BibitemShut {NoStop}%
\bibitem [{\citenamefont {Chen}\ \emph {et~al.}(1999)\citenamefont {Chen},
  \citenamefont {Lov{\'a}sz},\ and\ \citenamefont {Pak}}]{chen1999lifting}%
  \BibitemOpen
  \bibfield  {author} {\bibinfo {author} {\bibfnamefont {F.}~\bibnamefont
  {Chen}}, \bibinfo {author} {\bibfnamefont {L.}~\bibnamefont {Lov{\'a}sz}},\
  and\ \bibinfo {author} {\bibfnamefont {I.}~\bibnamefont {Pak}},\ }\bibfield
  {title} {\bibinfo {title} {Lifting markov chains to speed up mixing},\ }in\
  \href@noop {} {\emph {\bibinfo {booktitle} {Proc. 31st Ann. ACM Symp. Theory
  of Comp.}}}\ (\bibinfo {year} {1999})\ pp.\ \bibinfo {pages}
  {275--281}\BibitemShut {NoStop}%
\bibitem [{\citenamefont {Diaconis}\ \emph {et~al.}(2000)\citenamefont
  {Diaconis}, \citenamefont {Holmes},\ and\ \citenamefont
  {Neal}}]{diaconis2000analysis}%
  \BibitemOpen
  \bibfield  {author} {\bibinfo {author} {\bibfnamefont {P.}~\bibnamefont
  {Diaconis}}, \bibinfo {author} {\bibfnamefont {S.}~\bibnamefont {Holmes}},\
  and\ \bibinfo {author} {\bibfnamefont {R.~M.}\ \bibnamefont {Neal}},\
  }\bibfield  {title} {\bibinfo {title} {Analysis of a nonreversible markov
  chain sampler},\ }\href@noop {} {\bibfield  {journal} {\bibinfo  {journal}
  {Annals of Applied Probability}\ ,\ \bibinfo {pages} {726}} (\bibinfo {year}
  {2000})}\BibitemShut {NoStop}%
\bibitem [{\citenamefont {Scalliet}\ \emph {et~al.}(2022)\citenamefont
  {Scalliet}, \citenamefont {Guiselin},\ and\ \citenamefont
  {Berthier}}]{scalliet2022thirty}%
  \BibitemOpen
  \bibfield  {author} {\bibinfo {author} {\bibfnamefont {C.}~\bibnamefont
  {Scalliet}}, \bibinfo {author} {\bibfnamefont {B.}~\bibnamefont {Guiselin}},\
  and\ \bibinfo {author} {\bibfnamefont {L.}~\bibnamefont {Berthier}},\
  }\bibfield  {title} {\bibinfo {title} {Thirty milliseconds in the life of a
  supercooled liquid},\ }\href@noop {} {\bibfield  {journal} {\bibinfo
  {journal} {Physical Review X}\ }\textbf {\bibinfo {volume} {12}},\ \bibinfo
  {pages} {041028} (\bibinfo {year} {2022})}\BibitemShut {NoStop}%
\bibitem [{\citenamefont {Bouchaud}(2024)}]{bouchaud2024dynamics}%
  \BibitemOpen
  \bibfield  {author} {\bibinfo {author} {\bibfnamefont {J.-P.}\ \bibnamefont
  {Bouchaud}},\ }\bibfield  {title} {\bibinfo {title} {Why is the dynamics of
  glasses super-arrhenius?},\ }\href@noop {} {\bibfield  {journal} {\bibinfo
  {journal} {arXiv preprint arXiv:2402.01883}\ } (\bibinfo {year}
  {2024})}\BibitemShut {NoStop}%
\bibitem [{\citenamefont {Charbonneau}\ \emph {et~al.}(2017)\citenamefont
  {Charbonneau}, \citenamefont {Kurchan}, \citenamefont {Parisi}, \citenamefont
  {Urbani},\ and\ \citenamefont {Zamponi}}]{charbonneau2017glass}%
  \BibitemOpen
  \bibfield  {author} {\bibinfo {author} {\bibfnamefont {P.}~\bibnamefont
  {Charbonneau}}, \bibinfo {author} {\bibfnamefont {J.}~\bibnamefont
  {Kurchan}}, \bibinfo {author} {\bibfnamefont {G.}~\bibnamefont {Parisi}},
  \bibinfo {author} {\bibfnamefont {P.}~\bibnamefont {Urbani}},\ and\ \bibinfo
  {author} {\bibfnamefont {F.}~\bibnamefont {Zamponi}},\ }\bibfield  {title}
  {\bibinfo {title} {Glass and jamming transitions: From exact results to
  finite-dimensional descriptions},\ }\href@noop {} {\bibfield  {journal}
  {\bibinfo  {journal} {Annual Review of Condensed Matter Physics}\ }\textbf
  {\bibinfo {volume} {8}},\ \bibinfo {pages} {265} (\bibinfo {year}
  {2017})}\BibitemShut {NoStop}%
\bibitem [{\citenamefont {Parisi}\ \emph {et~al.}(2020)\citenamefont {Parisi},
  \citenamefont {Urbani},\ and\ \citenamefont {Zamponi}}]{parisi2020theory}%
  \BibitemOpen
  \bibfield  {author} {\bibinfo {author} {\bibfnamefont {G.}~\bibnamefont
  {Parisi}}, \bibinfo {author} {\bibfnamefont {P.}~\bibnamefont {Urbani}},\
  and\ \bibinfo {author} {\bibfnamefont {F.}~\bibnamefont {Zamponi}},\
  }\href@noop {} {\emph {\bibinfo {title} {Theory of simple glasses}}}\
  (\bibinfo  {publisher} {Cambridge University Press},\ \bibinfo {year}
  {2020})\BibitemShut {NoStop}%
\bibitem [{\citenamefont {Frisch}\ \emph {et~al.}(1985)\citenamefont {Frisch},
  \citenamefont {Rivier},\ and\ \citenamefont {Wyler}}]{frisch1985classical}%
  \BibitemOpen
  \bibfield  {author} {\bibinfo {author} {\bibfnamefont {H.}~\bibnamefont
  {Frisch}}, \bibinfo {author} {\bibfnamefont {N.}~\bibnamefont {Rivier}},\
  and\ \bibinfo {author} {\bibfnamefont {D.}~\bibnamefont {Wyler}},\ }\bibfield
   {title} {\bibinfo {title} {Classical hard-sphere fluid in infinitely many
  dimensions},\ }\href@noop {} {\bibfield  {journal} {\bibinfo  {journal}
  {Physical review letters}\ }\textbf {\bibinfo {volume} {54}},\ \bibinfo
  {pages} {2061} (\bibinfo {year} {1985})}\BibitemShut {NoStop}%
\bibitem [{\citenamefont {Kurchan}\ \emph {et~al.}(2012)\citenamefont
  {Kurchan}, \citenamefont {Parisi},\ and\ \citenamefont
  {Zamponi}}]{kurchan2012exact}%
  \BibitemOpen
  \bibfield  {author} {\bibinfo {author} {\bibfnamefont {J.}~\bibnamefont
  {Kurchan}}, \bibinfo {author} {\bibfnamefont {G.}~\bibnamefont {Parisi}},\
  and\ \bibinfo {author} {\bibfnamefont {F.}~\bibnamefont {Zamponi}},\
  }\bibfield  {title} {\bibinfo {title} {Exact theory of dense amorphous hard
  spheres in high dimension i. the free energy},\ }\href@noop {} {\bibfield
  {journal} {\bibinfo  {journal} {Journal of Statistical Mechanics: Theory and
  Experiment}\ }\textbf {\bibinfo {volume} {2012}},\ \bibinfo {pages} {P10012}
  (\bibinfo {year} {2012})}\BibitemShut {NoStop}%
\bibitem [{\citenamefont {Maimbourg}\ \emph {et~al.}(2016)\citenamefont
  {Maimbourg}, \citenamefont {Kurchan},\ and\ \citenamefont
  {Zamponi}}]{maimbourg2016solution}%
  \BibitemOpen
  \bibfield  {author} {\bibinfo {author} {\bibfnamefont {T.}~\bibnamefont
  {Maimbourg}}, \bibinfo {author} {\bibfnamefont {J.}~\bibnamefont {Kurchan}},\
  and\ \bibinfo {author} {\bibfnamefont {F.}~\bibnamefont {Zamponi}},\
  }\bibfield  {title} {\bibinfo {title} {Solution of the dynamics of liquids in
  the large-dimensional limit},\ }\href@noop {} {\bibfield  {journal} {\bibinfo
   {journal} {Physical review letters}\ }\textbf {\bibinfo {volume} {116}},\
  \bibinfo {pages} {015902} (\bibinfo {year} {2016})}\BibitemShut {NoStop}%
\bibitem [{\citenamefont {Agoritsas}\ \emph
  {et~al.}(2019{\natexlab{a}})\citenamefont {Agoritsas}, \citenamefont
  {Maimbourg},\ and\ \citenamefont {Zamponi}}]{agoritsas2019out1}%
  \BibitemOpen
  \bibfield  {author} {\bibinfo {author} {\bibfnamefont {E.}~\bibnamefont
  {Agoritsas}}, \bibinfo {author} {\bibfnamefont {T.}~\bibnamefont
  {Maimbourg}},\ and\ \bibinfo {author} {\bibfnamefont {F.}~\bibnamefont
  {Zamponi}},\ }\bibfield  {title} {\bibinfo {title} {Out-of-equilibrium
  dynamical equations of infinite-dimensional particle systems i. the isotropic
  case},\ }\href@noop {} {\bibfield  {journal} {\bibinfo  {journal} {Journal of
  Physics A: Mathematical and Theoretical}\ }\textbf {\bibinfo {volume} {52}},\
  \bibinfo {pages} {144002} (\bibinfo {year} {2019}{\natexlab{a}})}\BibitemShut
  {NoStop}%
\bibitem [{\citenamefont {Agoritsas}\ \emph
  {et~al.}(2019{\natexlab{b}})\citenamefont {Agoritsas}, \citenamefont
  {Maimbourg},\ and\ \citenamefont {Zamponi}}]{agoritsas2019out2}%
  \BibitemOpen
  \bibfield  {author} {\bibinfo {author} {\bibfnamefont {E.}~\bibnamefont
  {Agoritsas}}, \bibinfo {author} {\bibfnamefont {T.}~\bibnamefont
  {Maimbourg}},\ and\ \bibinfo {author} {\bibfnamefont {F.}~\bibnamefont
  {Zamponi}},\ }\bibfield  {title} {\bibinfo {title} {Out-of-equilibrium
  dynamical equations of infinite-dimensional particle systems. ii. the
  anisotropic case under shear strain},\ }\href@noop {} {\bibfield  {journal}
  {\bibinfo  {journal} {Journal of Physics A: Mathematical and Theoretical}\
  }\textbf {\bibinfo {volume} {52}},\ \bibinfo {pages} {334001} (\bibinfo
  {year} {2019}{\natexlab{b}})}\BibitemShut {NoStop}%
\bibitem [{\citenamefont {Liu}\ \emph {et~al.}(2021)\citenamefont {Liu},
  \citenamefont {Biroli}, \citenamefont {Reichman},\ and\ \citenamefont
  {Szamel}}]{liu2021dynamics}%
  \BibitemOpen
  \bibfield  {author} {\bibinfo {author} {\bibfnamefont {C.}~\bibnamefont
  {Liu}}, \bibinfo {author} {\bibfnamefont {G.}~\bibnamefont {Biroli}},
  \bibinfo {author} {\bibfnamefont {D.~R.}\ \bibnamefont {Reichman}},\ and\
  \bibinfo {author} {\bibfnamefont {G.}~\bibnamefont {Szamel}},\ }\bibfield
  {title} {\bibinfo {title} {Dynamics of liquids in the large-dimensional
  limit},\ }\href@noop {} {\bibfield  {journal} {\bibinfo  {journal} {Physical
  Review E}\ }\textbf {\bibinfo {volume} {104}},\ \bibinfo {pages} {054606}
  (\bibinfo {year} {2021})}\BibitemShut {NoStop}%
\bibitem [{\citenamefont {Biroli}\ \emph {et~al.}(2022)\citenamefont {Biroli},
  \citenamefont {Charbonneau}, \citenamefont {Folena}, \citenamefont {Hu},\
  and\ \citenamefont {Zamponi}}]{biroli2022local}%
  \BibitemOpen
  \bibfield  {author} {\bibinfo {author} {\bibfnamefont {G.}~\bibnamefont
  {Biroli}}, \bibinfo {author} {\bibfnamefont {P.}~\bibnamefont {Charbonneau}},
  \bibinfo {author} {\bibfnamefont {G.}~\bibnamefont {Folena}}, \bibinfo
  {author} {\bibfnamefont {Y.}~\bibnamefont {Hu}},\ and\ \bibinfo {author}
  {\bibfnamefont {F.}~\bibnamefont {Zamponi}},\ }\bibfield  {title} {\bibinfo
  {title} {Local dynamical heterogeneity in simple glass formers},\ }\href@noop
  {} {\bibfield  {journal} {\bibinfo  {journal} {Physical review letters}\
  }\textbf {\bibinfo {volume} {128}},\ \bibinfo {pages} {175501} (\bibinfo
  {year} {2022})}\BibitemShut {NoStop}%
\bibitem [{\citenamefont {M{\'e}zard}\ and\ \citenamefont
  {Parisi}(1999)}]{mezard1999thermodynamics}%
  \BibitemOpen
  \bibfield  {author} {\bibinfo {author} {\bibfnamefont {M.}~\bibnamefont
  {M{\'e}zard}}\ and\ \bibinfo {author} {\bibfnamefont {G.}~\bibnamefont
  {Parisi}},\ }\bibfield  {title} {\bibinfo {title} {Thermodynamics of glasses:
  A first principles computation},\ }\href@noop {} {\bibfield  {journal}
  {\bibinfo  {journal} {Journal of Physics: Condensed Matter}\ }\textbf
  {\bibinfo {volume} {11}},\ \bibinfo {pages} {A157} (\bibinfo {year}
  {1999})}\BibitemShut {NoStop}%
\bibitem [{\citenamefont {Parisi}\ and\ \citenamefont
  {Zamponi}(2010)}]{parisi2010mean}%
  \BibitemOpen
  \bibfield  {author} {\bibinfo {author} {\bibfnamefont {G.}~\bibnamefont
  {Parisi}}\ and\ \bibinfo {author} {\bibfnamefont {F.}~\bibnamefont
  {Zamponi}},\ }\bibfield  {title} {\bibinfo {title} {Mean-field theory of hard
  sphere glasses and jamming},\ }\href@noop {} {\bibfield  {journal} {\bibinfo
  {journal} {Reviews of Modern Physics}\ }\textbf {\bibinfo {volume} {82}},\
  \bibinfo {pages} {789} (\bibinfo {year} {2010})}\BibitemShut {NoStop}%
\bibitem [{\citenamefont {Monasson}(1995)}]{monasson1995structural}%
  \BibitemOpen
  \bibfield  {author} {\bibinfo {author} {\bibfnamefont {R.}~\bibnamefont
  {Monasson}},\ }\bibfield  {title} {\bibinfo {title} {Structural glass
  transition and the entropy of the metastable states},\ }\href@noop {}
  {\bibfield  {journal} {\bibinfo  {journal} {Physical review letters}\
  }\textbf {\bibinfo {volume} {75}},\ \bibinfo {pages} {2847} (\bibinfo {year}
  {1995})}\BibitemShut {NoStop}%
\bibitem [{\citenamefont {Franz}\ and\ \citenamefont
  {Parisi}(1997)}]{franz1997phase}%
  \BibitemOpen
  \bibfield  {author} {\bibinfo {author} {\bibfnamefont {S.}~\bibnamefont
  {Franz}}\ and\ \bibinfo {author} {\bibfnamefont {G.}~\bibnamefont {Parisi}},\
  }\bibfield  {title} {\bibinfo {title} {Phase diagram of coupled glassy
  systems: A mean-field study},\ }\href@noop {} {\bibfield  {journal} {\bibinfo
   {journal} {Physical review letters}\ }\textbf {\bibinfo {volume} {79}},\
  \bibinfo {pages} {2486} (\bibinfo {year} {1997})}\BibitemShut {NoStop}%
\bibitem [{\citenamefont {Futami}\ \emph {et~al.}(2020)\citenamefont {Futami},
  \citenamefont {Sato},\ and\ \citenamefont
  {Sugiyama}}]{futami2020accelerating}%
  \BibitemOpen
  \bibfield  {author} {\bibinfo {author} {\bibfnamefont {F.}~\bibnamefont
  {Futami}}, \bibinfo {author} {\bibfnamefont {I.}~\bibnamefont {Sato}},\ and\
  \bibinfo {author} {\bibfnamefont {M.}~\bibnamefont {Sugiyama}},\ }\bibfield
  {title} {\bibinfo {title} {Accelerating the diffusion-based ensemble sampling
  by non-reversible dynamics},\ }in\ \href@noop {} {\emph {\bibinfo {booktitle}
  {International Conference on Machine Learning}}}\ (\bibinfo {organization}
  {PMLR},\ \bibinfo {year} {2020})\ pp.\ \bibinfo {pages}
  {3337--3347}\BibitemShut {NoStop}%
\bibitem [{\citenamefont {Gao}\ \emph {et~al.}(2020)\citenamefont {Gao},
  \citenamefont {Gurbuzbalaban},\ and\ \citenamefont {Zhu}}]{gao2020breaking}%
  \BibitemOpen
  \bibfield  {author} {\bibinfo {author} {\bibfnamefont {X.}~\bibnamefont
  {Gao}}, \bibinfo {author} {\bibfnamefont {M.}~\bibnamefont {Gurbuzbalaban}},\
  and\ \bibinfo {author} {\bibfnamefont {L.}~\bibnamefont {Zhu}},\ }\bibfield
  {title} {\bibinfo {title} {Breaking reversibility accelerates langevin
  dynamics for non-convex optimization},\ }\href@noop {} {\bibfield  {journal}
  {\bibinfo  {journal} {Advances in Neural Information Processing Systems}\
  }\textbf {\bibinfo {volume} {33}},\ \bibinfo {pages} {17850} (\bibinfo {year}
  {2020})}\BibitemShut {NoStop}%
\bibitem [{\citenamefont {Ghimenti}\ \emph {et~al.}(2023)\citenamefont
  {Ghimenti}, \citenamefont {Berthier}, \citenamefont {Szamel},\ and\
  \citenamefont {van Wijland}}]{ghimenti2023sampling}%
  \BibitemOpen
  \bibfield  {author} {\bibinfo {author} {\bibfnamefont {F.}~\bibnamefont
  {Ghimenti}}, \bibinfo {author} {\bibfnamefont {L.}~\bibnamefont {Berthier}},
  \bibinfo {author} {\bibfnamefont {G.}~\bibnamefont {Szamel}},\ and\ \bibinfo
  {author} {\bibfnamefont {F.}~\bibnamefont {van Wijland}},\ }\bibfield
  {title} {\bibinfo {title} {Sampling efficiency of transverse forces in dense
  liquids},\ }\href@noop {} {\bibfield  {journal} {\bibinfo  {journal}
  {Physical Review Letters}\ }\textbf {\bibinfo {volume} {131}},\ \bibinfo
  {pages} {257101} (\bibinfo {year} {2023})}\BibitemShut {NoStop}%
\bibitem [{\citenamefont {Ghimenti}\ and\ \citenamefont {van
  Wijland}(2022)}]{ghimenti2022accelerating}%
  \BibitemOpen
  \bibfield  {author} {\bibinfo {author} {\bibfnamefont {F.}~\bibnamefont
  {Ghimenti}}\ and\ \bibinfo {author} {\bibfnamefont {F.}~\bibnamefont {van
  Wijland}},\ }\bibfield  {title} {\bibinfo {title} {Accelerating, to some
  extent, the $p$-spin dynamics},\ }\href
  {https://doi.org/10.1103/PhysRevE.105.054137} {\bibfield  {journal} {\bibinfo
   {journal} {Phys. Rev. E}\ }\textbf {\bibinfo {volume} {105}},\ \bibinfo
  {pages} {054137} (\bibinfo {year} {2022})}\BibitemShut {NoStop}%
\bibitem [{\citenamefont {Ikeda}\ \emph {et~al.}(2017)\citenamefont {Ikeda},
  \citenamefont {Zamponi},\ and\ \citenamefont {Ikeda}}]{ikeda2017mean}%
  \BibitemOpen
  \bibfield  {author} {\bibinfo {author} {\bibfnamefont {H.}~\bibnamefont
  {Ikeda}}, \bibinfo {author} {\bibfnamefont {F.}~\bibnamefont {Zamponi}},\
  and\ \bibinfo {author} {\bibfnamefont {A.}~\bibnamefont {Ikeda}},\ }\bibfield
   {title} {\bibinfo {title} {Mean field theory of the swap monte carlo
  algorithm},\ }\href@noop {} {\bibfield  {journal} {\bibinfo  {journal} {The
  Journal of chemical physics}\ }\textbf {\bibinfo {volume} {147}} (\bibinfo
  {year} {2017})}\BibitemShut {NoStop}%
\bibitem [{\citenamefont {Szamel}(2018)}]{szamel2018theory}%
  \BibitemOpen
  \bibfield  {author} {\bibinfo {author} {\bibfnamefont {G.}~\bibnamefont
  {Szamel}},\ }\bibfield  {title} {\bibinfo {title} {Theory for the dynamics of
  glassy mixtures with particle size swaps},\ }\href@noop {} {\bibfield
  {journal} {\bibinfo  {journal} {Physical Review E}\ }\textbf {\bibinfo
  {volume} {98}},\ \bibinfo {pages} {050601} (\bibinfo {year}
  {2018})}\BibitemShut {NoStop}%
\bibitem [{\citenamefont {Hargus}\ \emph {et~al.}(2021)\citenamefont {Hargus},
  \citenamefont {Epstein},\ and\ \citenamefont {Mandadapu}}]{hargus2021odd}%
  \BibitemOpen
  \bibfield  {author} {\bibinfo {author} {\bibfnamefont {C.}~\bibnamefont
  {Hargus}}, \bibinfo {author} {\bibfnamefont {J.~M.}\ \bibnamefont
  {Epstein}},\ and\ \bibinfo {author} {\bibfnamefont {K.~K.}\ \bibnamefont
  {Mandadapu}},\ }\bibfield  {title} {\bibinfo {title} {Odd diffusivity of
  chiral random motion},\ }\href@noop {} {\bibfield  {journal} {\bibinfo
  {journal} {Physical review letters}\ }\textbf {\bibinfo {volume} {127}},\
  \bibinfo {pages} {178001} (\bibinfo {year} {2021})}\BibitemShut {NoStop}%
\bibitem [{\citenamefont {Poggioli}\ and\ \citenamefont
  {Limmer}(2023)}]{poggioli2023odd}%
  \BibitemOpen
  \bibfield  {author} {\bibinfo {author} {\bibfnamefont {A.~R.}\ \bibnamefont
  {Poggioli}}\ and\ \bibinfo {author} {\bibfnamefont {D.~T.}\ \bibnamefont
  {Limmer}},\ }\bibfield  {title} {\bibinfo {title} {Odd mobility of a passive
  tracer in a chiral active fluid},\ }\href@noop {} {\bibfield  {journal}
  {\bibinfo  {journal} {Physical Review Letters}\ }\textbf {\bibinfo {volume}
  {130}},\ \bibinfo {pages} {158201} (\bibinfo {year} {2023})}\BibitemShut
  {NoStop}%
\bibitem [{\citenamefont {Banerjee}\ \emph {et~al.}(2017)\citenamefont
  {Banerjee}, \citenamefont {Souslov}, \citenamefont {Abanov},\ and\
  \citenamefont {Vitelli}}]{banerjee2017odd}%
  \BibitemOpen
  \bibfield  {author} {\bibinfo {author} {\bibfnamefont {D.}~\bibnamefont
  {Banerjee}}, \bibinfo {author} {\bibfnamefont {A.}~\bibnamefont {Souslov}},
  \bibinfo {author} {\bibfnamefont {A.~G.}\ \bibnamefont {Abanov}},\ and\
  \bibinfo {author} {\bibfnamefont {V.}~\bibnamefont {Vitelli}},\ }\bibfield
  {title} {\bibinfo {title} {Odd viscosity in chiral active fluids},\
  }\href@noop {} {\bibfield  {journal} {\bibinfo  {journal} {Nature
  communications}\ }\textbf {\bibinfo {volume} {8}},\ \bibinfo {pages} {1573}
  (\bibinfo {year} {2017})}\BibitemShut {NoStop}%
\bibitem [{\citenamefont {Fruchart}\ \emph {et~al.}(2021)\citenamefont
  {Fruchart}, \citenamefont {Hanai}, \citenamefont {Littlewood},\ and\
  \citenamefont {Vitelli}}]{fruchart2021non}%
  \BibitemOpen
  \bibfield  {author} {\bibinfo {author} {\bibfnamefont {M.}~\bibnamefont
  {Fruchart}}, \bibinfo {author} {\bibfnamefont {R.}~\bibnamefont {Hanai}},
  \bibinfo {author} {\bibfnamefont {P.~B.}\ \bibnamefont {Littlewood}},\ and\
  \bibinfo {author} {\bibfnamefont {V.}~\bibnamefont {Vitelli}},\ }\bibfield
  {title} {\bibinfo {title} {Non-reciprocal phase transitions},\ }\href@noop {}
  {\bibfield  {journal} {\bibinfo  {journal} {Nature}\ }\textbf {\bibinfo
  {volume} {592}},\ \bibinfo {pages} {363} (\bibinfo {year}
  {2021})}\BibitemShut {NoStop}%
\bibitem [{\citenamefont {J~Evans}\ and\ \citenamefont
  {P~Morriss}(2007)}]{j2007statistical}%
  \BibitemOpen
  \bibfield  {author} {\bibinfo {author} {\bibfnamefont {D.}~\bibnamefont
  {J~Evans}}\ and\ \bibinfo {author} {\bibfnamefont {G.}~\bibnamefont
  {P~Morriss}},\ }\href@noop {} {\emph {\bibinfo {title} {Statistical mechanics
  of nonequilbrium liquids}}}\ (\bibinfo  {publisher} {ANU Press},\ \bibinfo
  {year} {2007})\BibitemShut {NoStop}%
\bibitem [{\citenamefont {Zwanzig}(1960)}]{zwanzig1960ensemble}%
  \BibitemOpen
  \bibfield  {author} {\bibinfo {author} {\bibfnamefont {R.}~\bibnamefont
  {Zwanzig}},\ }\bibfield  {title} {\bibinfo {title} {Ensemble method in the
  theory of irreversibility},\ }\href@noop {} {\bibfield  {journal} {\bibinfo
  {journal} {The Journal of Chemical Physics}\ }\textbf {\bibinfo {volume}
  {33}},\ \bibinfo {pages} {1338} (\bibinfo {year} {1960})}\BibitemShut
  {NoStop}%
\bibitem [{\citenamefont {Manacorda}\ \emph {et~al.}(2020)\citenamefont
  {Manacorda}, \citenamefont {Schehr},\ and\ \citenamefont
  {Zamponi}}]{manacorda2020numerical}%
  \BibitemOpen
  \bibfield  {author} {\bibinfo {author} {\bibfnamefont {A.}~\bibnamefont
  {Manacorda}}, \bibinfo {author} {\bibfnamefont {G.}~\bibnamefont {Schehr}},\
  and\ \bibinfo {author} {\bibfnamefont {F.}~\bibnamefont {Zamponi}},\
  }\bibfield  {title} {\bibinfo {title} {Numerical solution of the dynamical
  mean field theory of infinite-dimensional equilibrium liquids},\ }\href@noop
  {} {\bibfield  {journal} {\bibinfo  {journal} {The Journal of chemical
  physics}\ }\textbf {\bibinfo {volume} {152}},\ \bibinfo {pages} {164506}
  (\bibinfo {year} {2020})}\BibitemShut {NoStop}%
\bibitem [{\citenamefont {G{\"o}tze}(2009)}]{gotze2009complex}%
  \BibitemOpen
  \bibfield  {author} {\bibinfo {author} {\bibfnamefont {W.}~\bibnamefont
  {G{\"o}tze}},\ }\href@noop {} {\emph {\bibinfo {title} {Complex dynamics of
  glass-forming liquids}}},\ Vol.\ \bibinfo {volume} {143}\ (\bibinfo
  {publisher} {Oxford University Press on Demand},\ \bibinfo {year}
  {2009})\BibitemShut {NoStop}%
\bibitem [{\citenamefont {Pottier}(2011)}]{POTTIER20112863}%
  \BibitemOpen
  \bibfield  {author} {\bibinfo {author} {\bibfnamefont {N.}~\bibnamefont
  {Pottier}},\ }\bibfield  {title} {\bibinfo {title} {Relaxation time
  distributions for an anomalously diffusing particle},\ }\href
  {https://doi.org/https://doi.org/10.1016/j.physa.2011.03.029} {\bibfield
  {journal} {\bibinfo  {journal} {Physica A: Statistical Mechanics and its
  Applications}\ }\textbf {\bibinfo {volume} {390}},\ \bibinfo {pages} {2863}
  (\bibinfo {year} {2011})}\BibitemShut {NoStop}%
\bibitem [{\citenamefont {Cugliandolo}\ \emph {et~al.}(1997)\citenamefont
  {Cugliandolo}, \citenamefont {Kurchan},\ and\ \citenamefont
  {Peliti}}]{cugliandolo1997energy}%
  \BibitemOpen
  \bibfield  {author} {\bibinfo {author} {\bibfnamefont {L.~F.}\ \bibnamefont
  {Cugliandolo}}, \bibinfo {author} {\bibfnamefont {J.}~\bibnamefont
  {Kurchan}},\ and\ \bibinfo {author} {\bibfnamefont {L.}~\bibnamefont
  {Peliti}},\ }\bibfield  {title} {\bibinfo {title} {Energy flow, partial
  equilibration, and effective temperatures in systems with slow dynamics},\
  }\href@noop {} {\bibfield  {journal} {\bibinfo  {journal} {Physical Review
  E}\ }\textbf {\bibinfo {volume} {55}},\ \bibinfo {pages} {3898} (\bibinfo
  {year} {1997})}\BibitemShut {NoStop}%
\bibitem [{\citenamefont {Epstein}\ and\ \citenamefont
  {Mandadapu}(2020)}]{epstein2020time}%
  \BibitemOpen
  \bibfield  {author} {\bibinfo {author} {\bibfnamefont {J.~M.}\ \bibnamefont
  {Epstein}}\ and\ \bibinfo {author} {\bibfnamefont {K.~K.}\ \bibnamefont
  {Mandadapu}},\ }\bibfield  {title} {\bibinfo {title} {Time-reversal symmetry
  breaking in two-dimensional nonequilibrium viscous fluids},\ }\href@noop {}
  {\bibfield  {journal} {\bibinfo  {journal} {Physical review E}\ }\textbf
  {\bibinfo {volume} {101}},\ \bibinfo {pages} {052614} (\bibinfo {year}
  {2020})}\BibitemShut {NoStop}%
\bibitem [{\citenamefont {Leli{\`e}vre}\ \emph {et~al.}(2013)\citenamefont
  {Leli{\`e}vre}, \citenamefont {Nier},\ and\ \citenamefont
  {Pavliotis}}]{lelievre2013optimal}%
  \BibitemOpen
  \bibfield  {author} {\bibinfo {author} {\bibfnamefont {T.}~\bibnamefont
  {Leli{\`e}vre}}, \bibinfo {author} {\bibfnamefont {F.}~\bibnamefont {Nier}},\
  and\ \bibinfo {author} {\bibfnamefont {G.~A.}\ \bibnamefont {Pavliotis}},\
  }\bibfield  {title} {\bibinfo {title} {Optimal non-reversible linear drift
  for the convergence to equilibrium of a diffusion},\ }\href@noop {}
  {\bibfield  {journal} {\bibinfo  {journal} {Journal of Statistical Physics}\
  }\textbf {\bibinfo {volume} {152}},\ \bibinfo {pages} {237} (\bibinfo {year}
  {2013})}\BibitemShut {NoStop}%
\end{thebibliography}%

\end{document}